\documentclass[sigconf,nonacm,balance=false,10pt]{acmart-arxiv} %

\usepackage{xurl}
\usepackage{booktabs}
\usepackage[ruled]{algorithm2e}
\usepackage{algpseudocode}
\usepackage{enumerate}
\usepackage[english]{babel}
\usepackage{blindtext}
\usepackage[caption=false,font=footnotesize,subrefformat=parens]{subfig}
\usepackage{amsmath}

\usepackage{amssymb}
\usepackage{xspace}
\usepackage{multirow}
\usepackage{threeparttable}
\usepackage{float}
\usepackage{flafter}
\usepackage{comment}
\usepackage[skip=12pt]{caption}
\usepackage{graphicx}
\usepackage{tikz}
\usetikzlibrary{arrows.meta,positioning}
\usepackage{listings}
\usepackage{pifont}
\newcommand{\cmark}{\ding{51}}
\newcommand{\xmark}{\ding{55}}
\newcommand{\one}{\ding{182}\xspace}
\newcommand{\two}{\ding{183}\xspace}
\newcommand{\three}{\ding{184}\xspace}
\newcommand{\four}{\ding{185}\xspace}
\newcommand{\five}{\ding{186}\xspace}
\usepackage{subcaption}
\usepackage{booktabs}
\usepackage{makecell}
\usepackage{amssymb}
\usepackage{multirow}
\usepackage[table]{xcolor}
\usepackage{pifont}
\usepackage{array}
\usepackage{subfig}

\usepackage{xcolor}
\usepackage[most]{tcolorbox}
\tcbuselibrary{listings}

\newtcblisting[auto counter]{pylisting}[2][]{
    float=t,                                  %
    listing only,
    title={Listing \thetcbcounter: #2},       %
    colback=white,                              %
    colframe=black!20,                        %
    coltitle=black,                           %
    fonttitle=\bfseries\footnotesize,         %
    arc=3pt,                                  %
    boxrule=0.5pt,                            %
    left=2pt, right=2pt, top=2pt, bottom=2pt, %
    #1,
    listing options={
        language=Python,
        basicstyle=\ttfamily\scriptsize,
        keywordstyle=\color{blue}\bfseries,
        stringstyle=\color{orange},
        commentstyle=\color{green!50!black}\itshape,
        emph={
            load, describe, compose, build, train, auto_align, from_tensor,
            set_exp_config, pilot_subc_interp, axis_normalize, conjugate_with_ref_rx,
            magnitude, PCA, stft, load_preset, run, rfnet, stfnet, FFT,
            rx_conjugate, axis_norm, pilot_interp,to_image,read,evaluate, resnet18
        },
        emphstyle=\color{purple}\bfseries,
        emph={[2]octo,octosense},
        emphstyle={[2]\color{violet}\bfseries},
        breaklines=true,
        columns=fullflexible,
        showstringspaces=false
    }
}

\definecolor{codegreen}{rgb}{0,0.6,0}
\definecolor{codegray}{rgb}{0.5,0.5,0.5}
\definecolor{codeblue}{rgb}{0.0,0.0,0.8}
\definecolor{backcolour}{rgb}{0.98,0.98,0.98}

\lstdefinestyle{octostyle}{
    backgroundcolor=\color{white},   
    commentstyle=\color{codegreen}\itshape, %
    keywordstyle=\color{codeblue}\bfseries, %
    stringstyle=\color{orange},             %
    basicstyle=\ttfamily\scriptsize,      %
    breakatwhitespace=false,         
    breaklines=true,                        %
    captionpos=t,                           %
    keepspaces=true,                        %
    columns=fullflexible,                   %
    showspaces=false,                
    showstringspaces=false,
    showtabs=false,                  
    tabsize=4
}

\def\fig{Fig.\xspace}

\def\tab{Tab.\xspace}

\def\ie{{\textit{i.e.}\xspace}} 
\def\eg{{\textit{e.g.}\xspace}}

\def\etc{{\textit{etc.}\xspace}}

\newcommand{\head}[1]{{\noindent \textbf{#1:}}}
\newcommand{\term}[1]{{\textit{#1}}}

\graphicspath{{./figures/}}
\graphicspath{{./pdf/}}
\DeclareGraphicsExtensions{.pdf,.jpeg,.png}

\ifodd 0
\newcommand{\rev}[1]{{\color{blue}#1}} %
\newcommand{\revold}[1]{#1}
\newcommand{\com}[1]{\textbf{\color{red}(COMMENT: #1)}} %
\newcommand{\todo}[1]{\textbf{{\color{orange}(TODO: #1)}}}
\else
\newcommand{\rev}[1]{#1}
\newcommand{\revold}[1]{#1}
\newcommand{\com}[1]{}
\newcommand{\todo}[1]{}
\fi

\renewcommand\footnotetextcopyrightpermission[1]{} %
\setcopyright{none}

\acmDOI{}

\acmISBN{}

\copyrightyear{}

\acmPrice{}

\def\sysname{\textsc{OctoSense}\xspace}

\begin{document}

\title[\sysname]{\sysname: Building a Unified Ecosystem for Open-Source Wireless Sensing}

\author{Weiying Hou}
\affiliation{
  \institution{The University of Hong Kong}
  \city{Hong Kong SAR}
  \country{China}
}
\email{weiying.hou@connect.hku.hk}

\author{Xie Zhang}
\affiliation{
  \institution{The University of Hong Kong}
  \city{Hong Kong SAR}
  \country{China}
}
\email{zhangxie@connect.hku.hk}

\author{Chenshu Wu}
\affiliation{
  \institution{The University of Hong Kong}
  \city{Hong Kong SAR}
  \country{China}
}
\email{chenshu@cs.hku.hk}

\renewcommand{\shortauthors}{Hou, Zhang, and Wu}

\begin{abstract}
As AI enters the physical world, wireless sensing has emerged as a key modality for enabling non-intrusive physical intelligence. 
However, unlike the mature ecosystems of vision and language models, wireless sensing research remains highly fragmented. 
The field faces a growing ``reproducibility wall'' caused by heterogeneous data formats, non-interoperable processing pipelines, and the lack of standardized benchmarks. 
To dismantle this barrier, we present \sysname, a unified platform designed to propel wireless sensing toward an ``ImageNet-style'' research ecosystem. 
\sysname introduces a holistic framework that decouples high-level model logic from ad-hoc data nuances through three distinct components: 
a unified data abstraction for streamlined dataset access, standardized signal operators for efficient model development, and a rigorous benchmark engine for rapid and fair comparison. 
To support existing datasets and models, \sysname integrates a comprehensive suite of widely used datasets and models, reducing the effort required for data and model adaptation to
a few lines of declarative code. 
We demonstrate the efficacy of \sysname through a set of usage examples that reproduce representative models on popular datasets. 
By providing a foundational open-source infrastructure, \sysname enables community efforts to accumulate rather than fragment, paving the way for wireless sensing research that is interoperable, reusable, and reproducible by design. 

\end{abstract}

\maketitle

\section{Introduction}
\label{sec:introduction}

\begin{figure}[t]
    \includegraphics[width=0.4\textwidth]{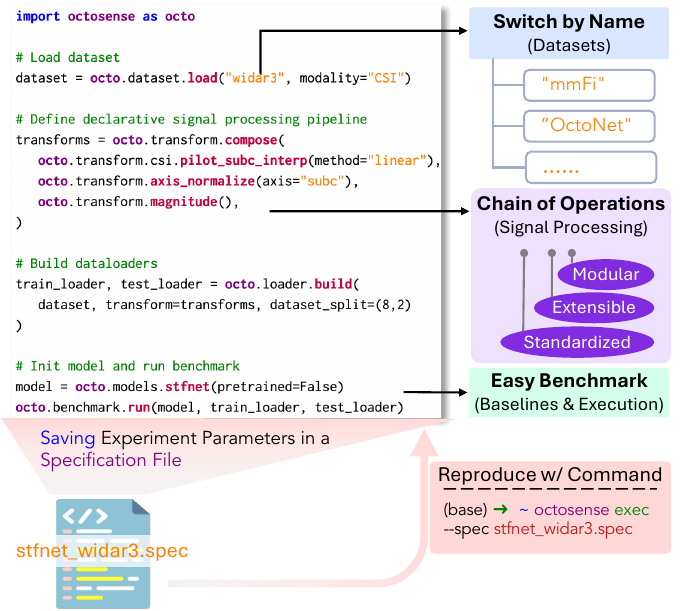}
    \vspace{-0.8\baselineskip}
\caption{\sysname overview through an STFNet-on-Widar3.0 example. 
\rm{\sysname organizes the conventional wireless sensing workflow into a unified interface that connects dataset selection, signal processing, dataloader construction, model invocation, and benchmark execution. Experimental settings are explicitly saved in a specification file, enabling the same baseline to be reproduced with a single command. This design reduces repeated engineering effort and enables wireless sensing research to be reused, compared, and extended with minimal effort.}}
    \label{fig:octosense_ov}
\end{figure}

As Artificial Intelligence (AI) transitions from the digital confines of large language models and vision foundation models into the physical world, the demand for multi-sensory physical intelligence has never been greater~\cite{liangVisionMultisensoryIntelligence2026}. 
Wireless sensing, \ie, sensing using various RF signals such as Wi-Fi, millimeter-wave (mmWave), UWB, 5G/6G, plays a critical role in this evolution, providing a non-intrusive, privacy-preserving, and through-the-wall lens into human activities~\cite{widar3,chi2024rf,csibench}, vital signs~\cite{zhengMoReFiMotionrobustFinegrained2021}, and human pose estimation~\cite{zhao2018rf}. 
Over the past decade, the field has witnessed an explosion of sophisticated algorithms and advanced models, fostering a range of applications from smart homes to healthcare to robotics~\cite{zhangWiDetectRobustMotion2019,yang2022artificial,lai2024enabling,boroushaki2021rfusion}. 

However, despite these technical milestones, the wireless sensing \term{ecosystem} remains in its infancy compared to the mature infrastructures of Computer Vision (CV) or Natural Language Processing (NLP). 
While a CV researcher can seamlessly benchmark a new model against ImageNet~\cite{deng2009imagenet} (or any major datasets) using standardized pipelines, the deep wireless sensing (DWS) landscape remains a fragmented collection of ``silos''~\cite{guarino2026survey}.
The lack of a sustainable ecosystem impedes unified \term{dataset access}, prevents interoperable \term{model development}, and renders rigorous \term{baseline benchmarking} nearly impossible. 
Without a unified platform to bridge these efforts, the field of wireless sensing risks getting trapped in a cycle of non-reusable resources, redundant engineering, and irreproducible results.

The root of this dilemma lies in the inherently heterogeneous and human-imperceptible nature of wireless signals, coupled with the absence of a common ``data language'' across the community. 
Unlike images or texts, which share standardized formats and are inherently perceptible by humans, wireless sensing data, ranging from Wi-Fi Channel State Information (CSI) to mmWave radar signals, are heterogeneous, hardware-dependent, and difficult to interpret, often stored in device- or user-specific formats with inconsistent sensing protocols and undocumented preprocessing steps~\cite{guarino2026survey}. 
The lack of a standardized, shared interface leads to what we term a ``reproducibility wall''. Currently, researchers are forced to rebuild dataset loaders and signal processing pipelines from scratch for almost every new design, making model development a highly repetitive process. Because of this isolated workflow, performance gains in one study are nearly impossible to verify fairly against another. When a baseline model fails to match its originally reported results, it is often not due to algorithmic flaws, but because of hidden discrepancies in processing parameters~\cite{widar3,Hou2024RFBoostUA}, dataset splits~\cite{varga2024mitigating,Kapoor2023LeakageAT}, and training configurations~\cite{Wightman2021ResNetSB}, \etc.
Critically, this systemic fragmentation cannot be solved simply by releasing more datasets or open-sourcing individual codes, as the burden of benchmarking across non-interoperable resources remains a significant barrier.
Furthermore, model re-implementation itself is notoriously difficult, involving many tedious and non-standard signal processing steps. 
To make matters worse, many critical parameter settings are hidden or omitted in the released implementations~\cite{guarino2026survey}, rendering accurate reproduction nearly impossible even when the original code is available.

To dismantle these barriers and foster a sustainable ecosystem, we introduce \sysname, the first platform to unify the fragmented landscape of wireless sensing research, shifting the paradigm from isolated development towards a mature engineering discipline. 
By providing a universal abstraction with a standardized nexus for datasets, models, and benchmarks, \sysname codifies the wireless sensing research cycle with a sustainable ecosystem, where heterogeneous datasets become interchangeable modules, diverse signal transforms become standard operators, and complex models become interoperable assets. 
\fig\ref{fig:octosense_ov} illustrates this idea using the reimplementation of STFNet \cite{yaoSTFNetsLearningSensing2019} on the Widar3.0 dataset \cite{widar3} for CSI-based gesture recognition. With \sysname, the end-to-end workflow can be expressed in about ten lines of code, whereas conventional practice often requires separate scripts for dataset loading, signal transformation, model reimplementation, and benchmark evaluation. Moreover, the full experimental configuration can be explicitly recorded in a specification file, \ie, \texttt{stfnet\_widar3.spec}, so that the same baseline can be reproduced through a single command in \sysname.
By abstracting away repeated dataset adaptation and workflow-specific engineering, \sysname allows researchers to focus on model development while remaining fully compatible with PyTorch~\cite{pytorch2019}. It also provides a shared interface for publishing, exposing, and reusing models through a model zoo, making it easier for subsequent work to build on prior implementations. 
Through an automated workflow, \sysname eliminates much of the burden of manual adaptation,  provides the community with a unified infrastructure through which collective effort can finally be compared, scaled, and sustained, with minimal manual efforts.

To achieve this, \sysname presents a holistic and systematic design where high-level model logic is decoupled from ad-hoc data formats and signal nuances, featuring three distinct components: 

\noindent\textbf{$\blacksquare$ Unified Data Abstraction}, which attaches explicit physical semantics, \eg, frequency, sampling rate, and antenna configurations, to raw heterogeneous data at the point of ingestion to enable seamless dataset access. 

\noindent\textbf{$\blacksquare$ Standardized Signal Operators}, which provide a modularized library of hardware-agnostic processing functions to facilitate interoperable model development. 

\noindent\textbf{$\blacksquare$ Rigorous Benchmarking Engine}, which establishes a specification-driven evaluation pipeline, enforcing exact and verifiable reproduction of results through a simple specification file to ensure fairness in baseline benchmarking.

With \sysname as a foundation, future research in wireless sensing becomes interoperable by design, ensuring that new models and datasets are immediately reusable and reproducible by the community.
To support existing workflows, \sysname has supported representative data processing pipelines (\eg, signal sanitation and spectrogram extraction), datasets 
including Widar3.0~\cite{widar3}, WiFi Gait~\cite{zhangGaitIDRobustWiFi2020}, SignFi~\cite{signfi}, FallDar~\cite{falldar}, CSI-Bench~\cite{csibench}, HuPR~\cite{hupr}, MM-Fi~\cite{mmfi}, XRF55~\cite{xrf55}, and OctoNet~\cite{octonet},
and mainstream models, including \revold{STFNet \cite{yaoSTFNetsLearningSensing2019}, UniTS \cite{UniTS}, RFNet \cite{dingRFnetUnifiedMetalearning2020}, Swin-Transformer \cite{liu2021swin}, ResNet \cite{resnet}}.
We validate the practical utility of \sysname through representative implementations, rigorously evaluating its architectural correctness, matching paired hand-engineered implementation accuracy within \revold{0.8\% points}, computational efficiency, and system effectiveness, reducing boilerplate code by over \revold{80\%}.
These capabilities are demonstrated through representative case studies, including \revold{out-of-the-box baseline inference, custom model benchmarking, declarative pipeline composition, and heterogeneous dataset integration}, showing how effortlessly our framework supports the full wireless sensing workflow.

\begin{figure}[t]
    \includegraphics[width=0.45\textwidth]{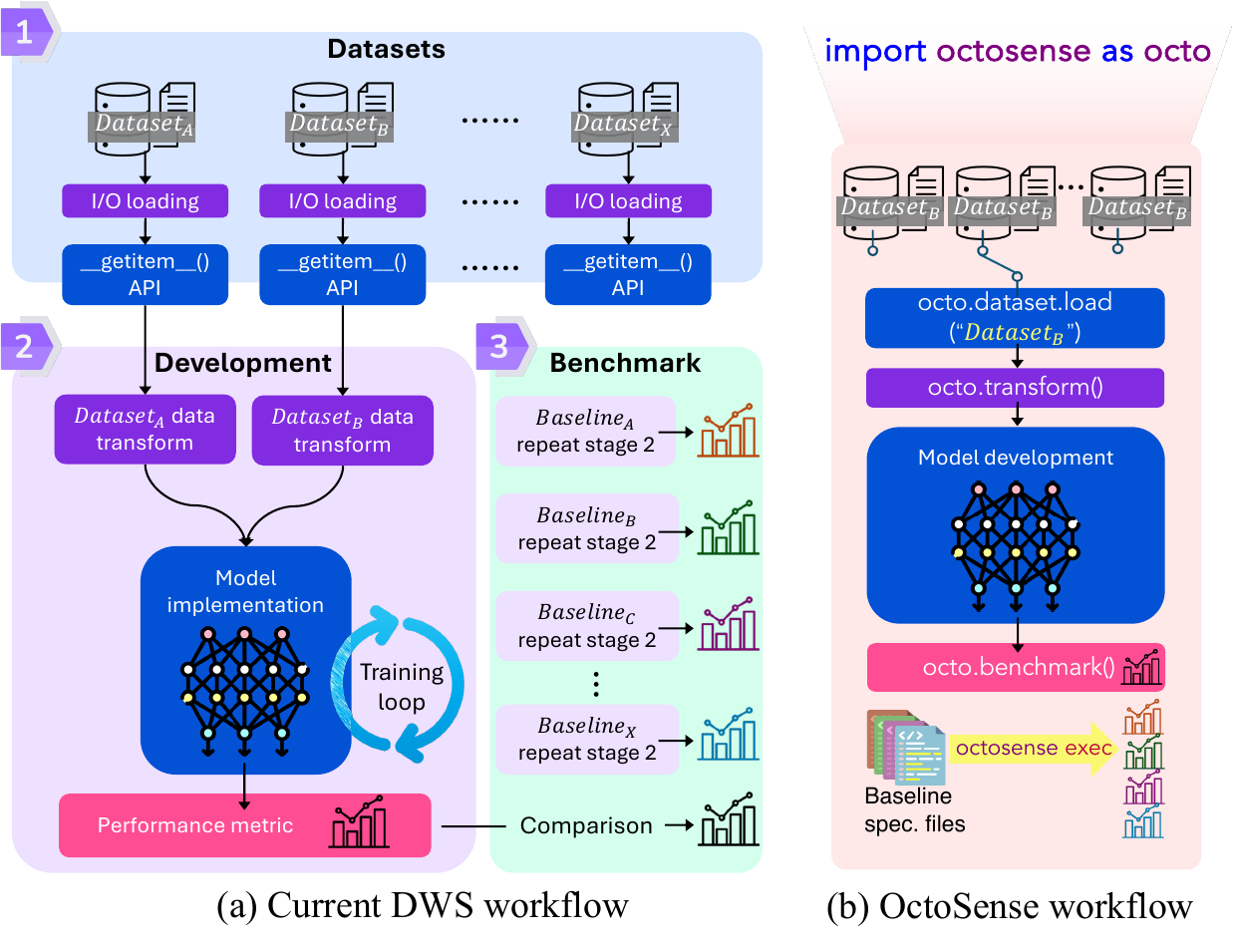}
    \vspace{-1.0\baselineskip}
\caption{Existing DWS research workflow and the \sysname-enabled workflow.  
\rm{(a) demonstrates a typical DWS study workflow involves three tightly coupled stages: dataset preparation, method development, and benchmarking. Although model implementation and training are supported by mature deep learning frameworks, the surrounding workflow remains highly dataset dependent and implementation specific, leading to repeated engineering effort and limited comparability across studies.
(b) \sysname reorganizes this workflow through a unified, semantic-aware interface that connects datasets, signal processing operators, models, and benchmark execution, thereby reducing repeated manual adaptation and improving reproducibility.
}}
    \label{fig:conventional_workflow}
\end{figure}

In summary, our core contribution is the design and implementation of \sysname, a unified and sustainable infrastructure to propel deep wireless sensing into the ``ImageNet era'', where datasets and models become reusable, reproducible, and easy to build upon. 
By lowering the barriers to dataset curation, model development, and cross-study comparison, \sysname aims to foster an ecosystem in which continuous community efforts accumulate rather than progress in isolation. 
We open-source \sysname at \revold{\url{https://github.com/aiot-lab/OctoSense}} and invite the community to contribute to its continued evolution. 
\section{Background and Motivation}
\label{sec:motivation}

In this section, we examine the current DWS research workflows, highlight the limitations that hinder efficient development, reproduction, and fair benchmarking, and review recurring issues in the open-source ecosystem and existing RF sensing frameworks.

\begin{figure}[t]
    \includegraphics[width=0.44\textwidth]{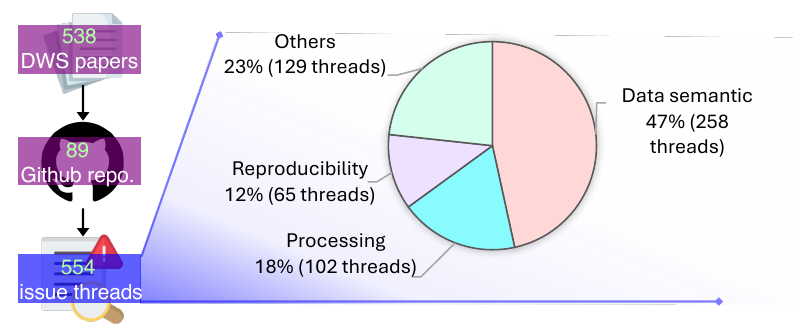}
    \vspace{-1.\baselineskip}
\caption{Open-source ecosystem survey of DWS research. 
\rm{Starting from a corpus of published DWS papers, we matched papers to public GitHub repositories and further examined community-side issue discussions. The resulting evidence chain reveals that the practical burden in current DWS research extends beyond model design itself, with recurrent discussions centered on data semantics, signal processing, and reproducibility. 
}}
    \label{fig:current_status_stats}
\end{figure}

\subsection{DWS Research Workflow}
\label{ssec:workflow}

As illustrated in \fig\ref{fig:conventional_workflow}(a), a conventional DWS study typically proceeds through three tightly coupled stages: dataset preparation, model development, and benchmarking. 
Although this workflow appears straightforward at a high level, each stage introduces significant fragmentation in practice, collectively making DWS research difficult to reuse, reproduce, and build upon.

\begin{table*}[t]
\centering
\footnotesize
\caption{Comparison of \sysname with representative RF sensing platforms across key functional dimensions. 
}
\vspace{-1.\baselineskip}
\label{tab:rf_solution_comparison}
\renewcommand{\arraystretch}{1.15}
\setlength{\tabcolsep}{6pt}
\begin{tabular}{lcccccc}
\toprule
\textbf{Frameworks } & 
\textbf{Modality} & 
\textbf{Unified Data API} & 
\textbf{Op./Model Lib.} & 
\textbf{Semantic Ops.} & 
\textbf{Spec-driven Benchmark} & 
\textbf{Reusable Pipeline} \\
\bottomrule
SenseFi~\cite{sensefi}                 & Wi-Fi           & \cmark & \cmark & \xmark & \xmark & \xmark \\
PicoScenes~\cite{picoscenes}              & Wi-Fi           & \cmark & \xmark & \xmark & \xmark & \xmark \\
CSI-Bench~\cite{csibench}               & Wi-Fi           & \cmark & \cmark & \xmark & \cmark & \xmark \\
OpenRadar~\cite{openradar2019}               & mmWave          & \cmark & \cmark & \xmark & \xmark & \xmark \\
\midrule
\textbf{\sysname}       & \textbf{RF} & \cmark & \cmark & \cmark & \cmark & \cmark \\
\bottomrule
\end{tabular}
\end{table*}

\head{Dataset Preparation}
The workflow begins with dataset preparation for a target sensing modality and application scenario.
Unlike standard CV benchmarks, wireless sensing datasets are rarely homogeneous in acquisition format or physical interpretation.
They are often collected with different hardware platforms, carrier frequencies, channel bandwidths, sampling rates, antenna configurations, and deployment layouts, while also varying in collection protocols such as subject orientation, environmental arrangement, and sensing distance.
As a result, the first step is not merely reading data from disk, but reconstructing signal representations suitable for learning. 
This process typically requires customized input/output routines, careful parsing of acquisition metadata, and explicit handling of signal axes, temporal organization, and storage efficiency. 
Existing tools may simplify hardware access and file decoding, but they usually stop at parser-level compatibility and do not preserve acquisition semantics through the workflow.
Consequently, datasets that appear structurally similar may still differ in physical meaning, label protocol, or preprocessing assumptions, making downstream reuse fragile from the outset.

Once the raw measurements are accessible, researchers must further define how the data are exposed to the learning framework, for example through dataset interfaces such as \texttt{\_\_getitem\_\_()} in PyTorch~\cite{pytorch2019}.
At this stage, raw measurements are converted into training instances, requiring decisions about tensor layout, sample granularity, label association, and metadata retention.
Consequently, the dataset interface implicitly encodes much of the signal interpretation logic that shapes all downstream training and evaluation.

\head{Method Development} Although neural network implementation and optimization are well supported by modern deep learning frameworks, DWS models typically rely on substantial signal processing that remains highly model-specific.  
In practice, DWS models are rarely trained directly on raw measurements. 
Instead, input signals are processed through a sequence of transformations, such as denoising, calibration, filtering, normalization, segmentation, Doppler or spectrogram construction, feature stacking, and tensor reshaping. 
These transformations depend jointly on sensing modality, acquisition setup, and model assumptions, yet they are often implemented as local scripts rather than reusable abstractions.
A concrete example can be found with Widar3 \cite{widar3}: Different studies reimplementing the CNN+GRU-based Widar3 model may adopt substantially different input representations, such as CSI-derived Doppler frequency spectrum (DFS) or body-coordinate velocity profile (BVP). Although these variants are often discussed under the same model name, they correspond to different signal representations and preprocessing assumptions, which can lead to significant differences in both performance and generalization \cite{Hou2024RFBoostUA}.

This creates two challenges. 
First, preprocessing protocols are often under-specified or scattered across code, comments, and supplementary materials, making it difficult to determine exact signal representation used in reported results. 
Second, even when a prior model is reimplemented faithfully, differences in preprocessing, sample construction, or tensor layout may silently alter the effective input distribution seen by the network.
In such cases, discrepancies in benchmark performance become difficult to attribute, because they may arise from architectural differences, signal representations, or mismatched assumptions introduced during reimplementation. 
The hidden processing layer between data loading and model execution therefore becomes a major source of inconsistency across studies. 

\head{Benchmarking}
The workflow typically concludes with benchmarking against prior methods. 
Researchers must often reproduce baseline models under identical datasets and evaluation settings to ensure fair comparison. 
In practice, however, many prior methods are not released with runnable code. Even when code is available, reproducing reported results still requires recovering undocumented details of dataset parsing, signal processing, training configuration, and evaluation protocols.
As a consequence, comparative evaluation in DWS frequently relies on repeated baseline reimplementation rather than direct reuse. 
Each new study may therefore reconstruct substantial parts of earlier workflows, including data interfaces, signal transforms, model wrappers, and metric computation, before a meaningful comparison can even begin.
This repeated effort slows research iteration and complicates results interpretation, as performance differences may partly reflect mismatched implementation assumptions rather than genuine algorithm advances.

In summary, these three forms of fragmentation lead not only to repeated engineering effort, but also to a persistent risk of silent misalignment in how wireless signals are parsed, transformed, and evaluated.
This observation motivates \sysname, shown in \fig\ref{fig:conventional_workflow}(b), which unifies the three stages through a semantic-aware workflow with name-based dataset selection, standardized signal processing transforms, and specification-driven benchmarking and reproduction.

\subsection{Current DWS Ecosystem}
\label{ssec:ecosystem}

\head{Survey on Open-source DWS Research}
To understand where the practical burden in current DWS research arises, we examine the open-source ecosystem associated with published wireless sensing studies.
Our goal was not to produce an exhaustive census of the field, but to construct an auditable evidence chain linking papers, released repositories, and community-side reproduction discussions.
As shown in \fig\ref{fig:current_status_stats}, starting from a broad corpus of wireless sensing papers, we matched published studies to public code repositories and then analyzed the corresponding GitHub issue threads.
After repository deduplication, we obtained 89 unique repositories, among which 61 contained issue discussions, yielding 554 issue threads for analysis.
We then used large language models to extract candidate evidence related to data semantics, preprocessing, and reproducibility, followed by manual verification of all retained labels. 
The detailed survey scope and verification procedure are provided in Appendix~\ref{app:survey-github-pipeline}.

Although these issue threads should not be interpreted as an unbiased prevalence estimate for the entire field, they provide direct evidence of the practical difficulties encountered by the community when attempting to understand datasets, reproduce signal processing procedures, and compare against prior methods.
Across the 554 manually reviewed threads, recurring discussions arise around data semantics, preprocessing, and reproducibility, indicating that the burden in DWS extends well beyond model design itself.
Instead, many difficulties stem from the workflow through which wireless signals are parsed, transformed, aligned, and evaluated.

\head{Existing RF Sensing Platforms} 
In recent years, the community has been accumulating more public datasets, including Widar3.0 \cite{widar3}, XRF55 \cite{xrf55} and XRF-V2 \cite{xrf_v2}, MM-Fi \cite{mmfi}, OctoNet \cite{octonet}, CSI-Bench \cite{csibench}, CUHK-X \cite{cuhkx}, HuPR\cite{hupr}, DeepSense~\cite{deepsense}, which largely mitigates the data scarcity issue in the field. A few sites are compiling available datasets \cite{sdp8,ieeedataport}. However, dataset accessibility remains notoriously troublesome, as discussed above. 
Besides datasets, efforts are also devoted to building benchmarks and data collection tools. 
For example, PicoScenes~\cite{picoscenes} facilitates data acquisition and parsing for WiFi sensing, OpenRadar~\cite{openradar2019} provides reusable components for mmWave signal processing, and benchmark-oriented platforms such as SenseFi~\cite{sensefi} and CSI-Bench~\cite{csibench} include a limited set of baseline models. 
As summarized in \tab\ref{tab:rf_solution_comparison}, existing efforts typically address only isolated parts of the workflow with limited coverage. 
Despite these advances, to the best of our knowledge, no prior platform unifies standardized dataset access, reusable signal processing and model development, and reproducible benchmarking.
Consequently, existing tools improve local usability but fail to resolve broader community fragmentation.

\section{\sysname Design}
\label{sec:design}

\sysname is a domain-specific platform built natively on top of PyTorch~\cite{pytorch2019} for deep wireless sensing. While PyTorch~\cite{pytorch2019} is highly successful for general deep learning, it treats input data as plain numeric arrays and does not naturally represent the physical semantics of heterogeneous wireless signals. By incorporating the physical characteristics of wireless signals into its core design, \sysname{} provides domain-optimized capabilities. 
With \sysname{}, the previously fragmented development process is transformed into a standardized and interoperable pipeline, allowing researchers to move away from building isolated scripts and instead handle complex RF data directly through their familiar PyTorch~\cite{pytorch2019} environment. 

\subsection{\sysname Overview}

\head{Dataset Access} Under the traditional paradigm, data collectors invent their own private file formats and provide undocumented metadata, which creates a huge parsing barrier for subsequent users. With \sysname{}, creators only need to write a simple driver script to adapt their raw hardware signals to our unified representation. The system automatically attaches the necessary semantics and specifications, making the newly collected dataset instantly readable and usable by any model in the ecosystem.

\head{Model Development} Currently, researchers write ad-hoc data loaders, manually figure out tensor shapes for each dataset by inspecting scattered documentation and source files, and hardcode the preprocessing steps. \sysname helps developers to quickly load datasets and switch between them by simply changing a configuration file. They can call a versioned preprocessing pipeline, read its clearly defined semantic output format, and seamlessly connect the processed data to their customized neural networks.

\head{Baseline Benchmarking} Previously, baselines were repeatedly reimplemented across different studies, and rigorous comparison was nearly impossible. With \sysname, preprocessing workflows are strictly bound to physical semantics rather than dynamic tensors. Researchers can reproduce a standard baseline with a single configuration file, ensuring fair and consistent evaluations across the entire community.

\begin{figure}[t]
    \includegraphics[width=0.45\textwidth]{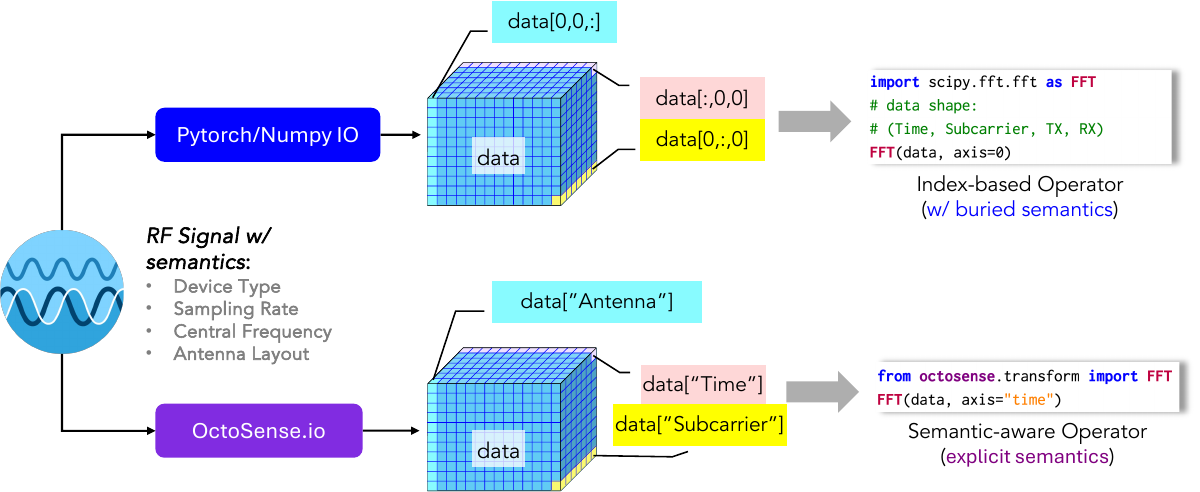}
    \vspace{-1\baselineskip}
\caption{Semantic-Aware Data Representation. 
\rm{Encoding RF semantics alongside data provides a user-friendly dataset interface while ensuring consistent signal interpretation and reducing potential careless preprocessing errors.
}}
    \label{fig:octosense_io}
\end{figure}

\subsection{Data Abstraction}
\label{ssec:implementation}

This subsection introduces the two abstractions that decouple downstream learning from device-specific RF formats. \texttt{\sysname{}.io} defines a semantic sample representation, while \texttt{\sysname{}.dataset} enables metadata-first operations over large collections of samples.

\head{$\blacksquare$ Unified Representation (\texttt{\sysname{}.io})} 
The core objective of this module is to convert raw wireless sensing data into a unified tensor representation equipped with explicit physical semantics. Unlike computer vision, where images share a standardized spatial structure, wireless sensing data exhibits substantial heterogeneity in dimensional organization and physical interpretation. In particular, we emphasize the concept of \emph{RF semantics}. For RF sensors, the same sensing semantic, such as time, antenna, or subcarrier, is often accompanied by device-dependent metadata. Moreover, such metadata is frequently processed separately from the signal tensor itself. In practice, this separation forces preprocessing pipelines to manipulate tensors through positional indices while managing metadata through auxiliary dictionaries or configuration files. As a result, the same semantic concept may correspond to different tensor axes or metadata fields across devices, making preprocessing logic tightly coupled to specific hardware formats and requiring individually customized parsing and transformation procedures. This fragmentation raises a key challenge: \textit{how to construct a unified representation that preserves RF semantics while decoupling downstream processing from device-specific tensor layouts and metadata formats?} To address this challenge, our IO design introduces two key mechanisms:

\head{\one Semantic-Aware Data Representation} Inspired by the data model of \texttt{xarray.DataArray}~\cite{xarray}, a Python library for coordinate-aware, labeled N-dimensional arrays, we represent wireless sensing data as a tensor whose dimensions are explicitly named using RF physical coordinates, such as time, subcarrier, antenna, and chirp, together with aligned metadata coordinates, as illustrated in \fig\ref{fig:octosense_io}. In this way, RF semantics are conveyed directly through dimension names rather than implicit tensor positions. This structure forms a semantically annotated tensor where physical meanings are attached to each axis. Depending on the task requirements, coordinates can optionally be preserved to retain physically meaningful axes, such as maintaining the time coordinate when studying temporal uniformity, while unnecessary coordinates can be discarded to balance computational efficiency.

\head{\two RF Semantic Mapping} To reconcile the heterogeneity across hardware platforms, we introduce a semantic mapping mechanism that converts device-specific metadata into a unified semantic schema. All downstream operations afterwards can interact exclusively with semantic labels, \eg, \texttt{data["time"]}, rather than fragile tensor axis indices, such as \texttt{data[0]}, or device-specific dictionary keys. As a result, the IO layer provides an extremely stable interface across datasets and devices, while metadata processing becomes naturally integrated with tensor operations, ensuring that signal data and associated metadata remain synchronized throughout the entire pipeline.

\begin{figure}[t]
    \includegraphics[width=0.42\textwidth]{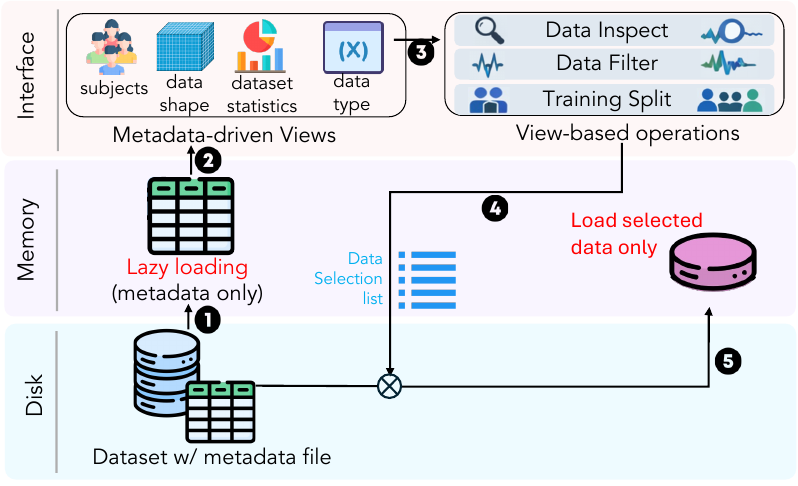}
    \vspace{-1.\baselineskip}
    \caption{Dataset management via metadata view-based operations in \texttt{\sysname{}.dataset}.\rm{
\one For large-scale RF datasets, only lightweight machine-readable metadata are first loaded into memory.
\two Based on this metadata, different dataset views are constructed without materializing the underlying RF data.
\three Users interact with these views to inspect the dataset and perform metadata-driven operations such as filtering and splitting.
\four Only the data selected through these operations are retained as the target subset for downstream use.
\five The corresponding RF tensors are loaded from disk only on demand, \ie, lazy loading.
}}
    \label{fig:lazy_loading}
\end{figure}

\head{$\blacksquare$ Metadata Management (\texttt{\sysname{}.dataset})} While the IO module provides a unified \emph{sample-level} representation, practical wireless sensing pipelines also require a \emph{dataset-level} abstraction for organizing, selecting, partitioning, and reusing large collections of samples. This need is especially critical for RF sensing, where datasets are often highly multidimensional and large-scale. If researchers must repeatedly go through an entire raw dataset merely to filter samples, inspect dimension-related properties, or construct train/validation/test splits, the resulting workflow becomes prohibitively inefficient. More fundamentally, these dataset-level operations depend primarily on metadata rather than on the raw signal tensor itself. However, metadata management in existing RF datasets is often fragmented and highly inconsistent. Important information is frequently buried in filename conventions, PDF annotations, scattered README files, ad-hoc toolboxes, or code comments, forcing researchers to manually infer how a dataset should be parsed and used. As a result, even simple questions—such as how many subjects performed a certain activity, or which samples belong to a specific environment or recording session—can require substantial manual effort. This fragmentation raises a key challenge: \textit{how to build a unified dataset abstraction that makes RF datasets both machine-readable and efficient to manipulate, while preserving the task-specific metadata needed by downstream learning pipelines?} To address this challenge, our dataset layer introduces two key mechanisms:

\head{\one Machine-Readable Metadata Management} Built upon the unified sample representation, \texttt{\sysname{}.dataset} introduces a standardized metadata layer that allows entire datasets to be uniformly understood, queried, managed, and reused by both developers and automated pipelines. Its most essential metadata is often the label, yet labels in wireless sensing are inherently \emph{task-dependent}: the same raw sample may correspond to different supervision targets for activity recognition, localization, gesture analysis, pose estimation, or other tasks. Moreover, labels can take heterogeneous forms, ranging from discrete categorical annotations (e.g., \texttt{sit}, \texttt{stand}) to structured targets such as keypoints, bounding boxes, or trajectories. To support this diversity, \sysname represents labels as metadata coordinates attached to the sample dimension. In this way, different task-specific labels can be managed in a unified format while remaining naturally aligned with the underlying samples. More generally, dataset metadata can be summarized and inspected through interfaces such as \texttt{dataset.describe()}, making dataset composition and supervision structure explicit before model training begins.

\head{\two View-Based Dataset Operations} Because RF datasets are large and high-dimensional, efficiency is a first-class requirement for dataset management. Inspired by the lazy loading design of \texttt{xarray.Dataset}, \sysname implements a \emph{view-based} operation mechanism for dataset-level manipulations, as depicted in \fig\ref{fig:lazy_loading}. Operations such as filtering, selection, and splitting create lightweight views over the original dataset instead of triggering large-scale data copies. As a result, heavy RF tensors remain lazily loaded, while most management logic is executed only on lightweight metadata. This design provides a convenient and efficient interface for development, making it practical to manipulate massive RF datasets without unnecessary memory or storage overhead. More importantly, by decoupling heavy signal storage from metadata-driven operations, \texttt{\sysname{}.dataset} enables a \emph{metadata-first} workflow: researchers can construct splits, select subsets, and plan preprocessing before materializing the actual signal tensors. This not only improves efficiency, but also opens the door to metadata-aware split validation, which can help reduce unintended data leakage caused by shared subjects, sessions, or environments across partitions.

\begin{figure}[t]
    \includegraphics[width=0.48\textwidth]{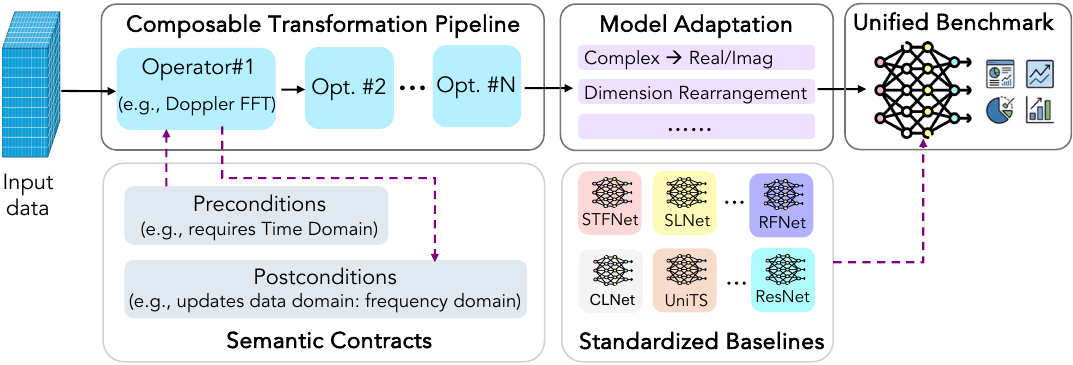}
    \vspace{-1.2\baselineskip}
\caption{Composable operators and standardized models. \rm{Input RF data are first processed through a composable transformation pipeline in \texttt{\sysname{}.transform}, where each operator is governed by semantic contracts that specify its preconditions and postconditions. These operators can be flexibly chained while preserving semantic correctness, and further adapt the processed data to model-specific input requirements. Built on top of this transform layer, \texttt{\sysname{}.models} provides a repository of standardized baseline architectures, reducing reimplementation effort and supporting consistent model usage across DWS tasks.}}
    \label{fig:operator_model}
\end{figure}

\subsection{Operators and Models}
To transform semantic RF data into model-ready inputs while preserving physical meaning, we propose contract-aware operators and a standardized model boundary (\fig~\ref{fig:operator_model}).

\head{$\blacksquare$ Semantic-Aware Operators (\texttt{\sysname{}.transform})} While \texttt{\sysname{}.io} and \texttt{\sysname{}.dataset} provide unified representations at the sample and dataset levels, wireless sensing pipelines still require a transformation layer that converts semantically structured RF data into model-ready inputs. This stage is particularly critical because signal preprocessing serves as the bridge between raw sensing data and downstream learning models. In practice, however, preprocessing in wireless sensing is often highly fragmented: researchers typically assemble ad-hoc sequences of signal processing functions, manually track metadata outside the tensor itself, and repeatedly rewrite format conversion logic for different sensors and models. Such fragmentation makes pipelines difficult to reuse, verify, and reproduce. More importantly, preprocessing errors are often semantic rather than syntactic. For example, an operator may silently assume a time-domain input while actually receiving a frequency-domain tensor, or reshape data in a way that breaks its physical interpretation. This raises a key challenge: \textit{how to build a composable operator system that remains flexible across heterogeneous RF modalities while preserving semantic correctness throughout the preprocessing pipeline?} To address this challenge, our transform layer introduces three key mechanisms:

\head{\one Composable Pipeline} Inspired by medical imaging frameworks such as MONAI~\cite{cardoso2022monai}, we adopt a dictionary-based transformation design. In this paradigm, data flows through the pipeline as a dictionary containing both signal tensors and their associated metadata, while each operator is implemented as a callable class that processes designated fields by key. This design naturally supports diverse input and output morphologies, including multi-channel WiFi CSI, mmWave heatmaps, and radar point clouds, without changing the core operator interface. As a result, preprocessing becomes modular and composable, rather than being tied to a specific signal format or model architecture.

\head{\two Semantic Contracts} To ensure semantic consistency during preprocessing, each operator is equipped with explicit semantic contracts that define its preconditions and postconditions through attributes such as \texttt{requires} and \texttt{updates}. For instance, a Doppler FFT operator can explicitly require its input to be in the time domain and, after execution, automatically update the metadata to indicate that the output is now in the frequency domain. This mechanism prevents silent logical errors caused by invalid operator ordering or incompatible input states, and allows preprocessing pipelines to behave in a predictable and verifiable manner. In this way, semantic information is no longer treated as an auxiliary byproduct, but becomes a first-class constraint of the transformation process itself.

\head{\three Model Adaptation} Beyond semantic correctness, operators must also adapt RF data to the practical requirements of downstream models. Wireless sensing signals often involve complex-valued representations, irregular tensor layouts, or modality-specific structures that cannot be directly consumed by common deep learning frameworks. To address this, \texttt{\sysname{}.transform} provides built-in support for model-oriented adaptation, including operations such as splitting complex values into real and imaginary components, extracting magnitude and phase, and rearranging dimensions to satisfy architecture-specific input conventions. Because each operator remains independent and semantically aware, these transformations can be chained into a unified preprocessing workflow. 
More importantly, the entire pipeline can be serialized into a declarative specification, enabling researchers to share, reproduce, and modify complex signal processing procedures with minimal overhead.

\head{$\blacksquare$ Model Repository (\texttt{\sysname{}.models})} A unified data and preprocessing ecosystem is only practically useful if it can connect directly to a standardized set of downstream models. In DWS research, however, a substantial amount of effort is repeatedly spent reimplementing existing architectures, adapting them to slightly different input formats, and reproducing baseline results under inconsistent training settings. This redundancy not only slows development, but also makes empirical comparisons unreliable. In many cases, differences in data formatting or interface assumptions are entangled with differences in model design, making it difficult to determine whether performance gains come from genuine architectural improvements or from mismatched experimental setups. This raises a key challenge: \textit{how to provide a unified modeling layer that reduces reimplementation burden while supporting fair and efficient benchmarking across sensing tasks?} To address this challenge, our model layer introduces two key mechanisms:

\head{\one Standardized Baselines} \texttt{\sysname{}.models} provides a curated repository of standardized baseline architectures covering representative neural network families used in wireless sensing research. We position this module not as a framework for inventing new architectures, but as a reliable foundation for benchmarking, reproduction, and comparison. By integrating these models into the unified data representation of \sysname, users can directly apply established baselines to their sensing tasks without rebuilding them from scratch. This substantially reduces redundant engineering effort and lowers the barrier to entry for both new practitioners and experienced researchers.

\head{\two Unified Benchmark Interface} Wireless sensing models often require different internal input layouts, and \sysname does not force them to share a single raw tensor shape. Instead, it standardizes the \emph{model boundary}: each model specifies its expected fields, layout, and semantic assumptions, while the transform layer adapts semantic RF data accordingly. Thus, the same RF sample can be converted into different model-specific inputs, such as image-like tensors for convolutional models or sequence-style tensors for transformers, without changing the surrounding training or evaluation code. Model outputs are likewise normalized into a shared task-level interface, enabling cleaner and fairer architectural comparison.

\begin{figure}[t]
\centering
    \includegraphics[width=0.48\textwidth]{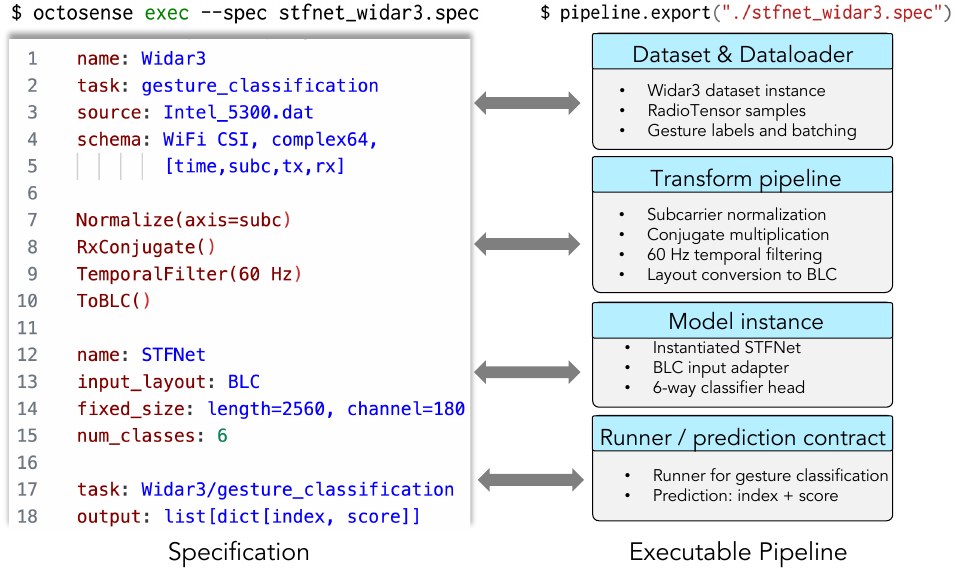}
    \vspace{-1.5\baselineskip}
\caption{Mutual conversion between benchmark specifications and executable pipelines. \rm{A benchmark can be instantiated from a declarative spec into an executable pipeline, and an executed pipeline can be exported back to the same spec for recording, restoration, and reuse.}}
\label{fig:spec_to_workflow}
\end{figure}

\subsection{Benchmark Engine}
Beyond unified data access and model integration, fair DWS benchmarking also requires experiments to be executed in a reusable way and recorded in a reproducible form. \sysname addresses this through a benchmark engine with two complementary layers: \texttt{pipelines} for benchmark execution and \texttt{specs} for benchmark serialization. In short, pipelines run a benchmark, whereas specs record and restore it, as illustrated in \fig~\ref{fig:spec_to_workflow}.

\head{$\blacksquare$ Reusable Workflows (\texttt{\sysname{}.pipelines})}
This module is the execution layer of the benchmark engine. A pipeline assembles dataset access, semantic transforms, model invocation, training, and evaluation into a reusable benchmark workflow, so that standard baselines can be launched without rewriting task-specific glue code. Because datasets, transforms, and models are modularized as interchangeable components, researchers can modify selected parts of a benchmark while keeping the rest of the evaluation protocol fixed, enabling cleaner and more efficient comparison.

\head{$\blacksquare$ Standardized Serialization (\texttt{\sysname{}.specs})}
The specification module is the serialization layer of the benchmark engine. A spec captures the full benchmark configuration, including datasets, transforms, models, and runtime parameters, as a unified and versionable description generated from the executed pipeline. This makes the specification file the minimal reproducible unit of benchmarking: it can be shared, reloaded, and rerun to recover the same experimental setup. By replacing scattered script arguments and manually maintained configuration files with a single structured record, \texttt{\sysname{}.specs} reduces configuration drift and improves long-term reproducibility.

\section{Implementation}
\label{sec:impl}
We implement \sysname{} as a modular PyTorch-based~\cite{pytorch2019} stack spanning raw signal IO, datasets, transforms, and benchmark models. At the IO layer, the current raw-signal readers already support major WiFi chipsets and capture formats, including IWL5300~\cite{5300}, Atheros~\cite{atherocsi}, ESP32~\cite{esp32}, Nexmon~\cite{nexmon}, and TI DCA1000~\cite{dca1000evm} for mmWave. On top of these readers, the current dataset layer covers representative benchmarks across multiple modalities, including Widar3.0~\cite{widar3}, WiFi Gait~\cite{zhangGaitIDRobustWiFi2020}, SignFi~\cite{signfi}, FallDar~\cite{falldar} and CSI-Bench~\cite{csibench} for WiFi, HuPR~\cite{hupr} for mmWave, and multimodal datasets such as MM-Fi~\cite{mmfi}, XRF55~\cite{xrf55}, and OctoNet~\cite{octonet}, which together support tasks including gesture and activity classification, pose estimation, and respiration monitoring. The transform layer provides the common processing components required by these workflows, including basic signal processing operators such as filtering, normalization and reusable preprocessing, as well as cuFFT-based~\cite{cufft} spectrum generation. The model layer further supports representative baseline architectures used in the current workflows, including STFNet~\cite{yaoSTFNetsLearningSensing2019}, UniTS~\cite{UniTS}, RFNet~\cite{dingRFnetUnifiedMetalearning2020}, Widar3~\cite{widar3}, Swin-Transformer \cite{liu2021swin}, and ResNet~\cite{resnet}. Taken together, the current implementation is organized around reusable reader-dataset-transform-model combinations, rather than isolated benchmark-specific scripts.

\begin{figure}[t]
    \centering
    \subfloat[STFNet $\times$ WiFi Gait\label{fig:gait_stfnet_loss}]{
        \includegraphics[width=0.22\textwidth]{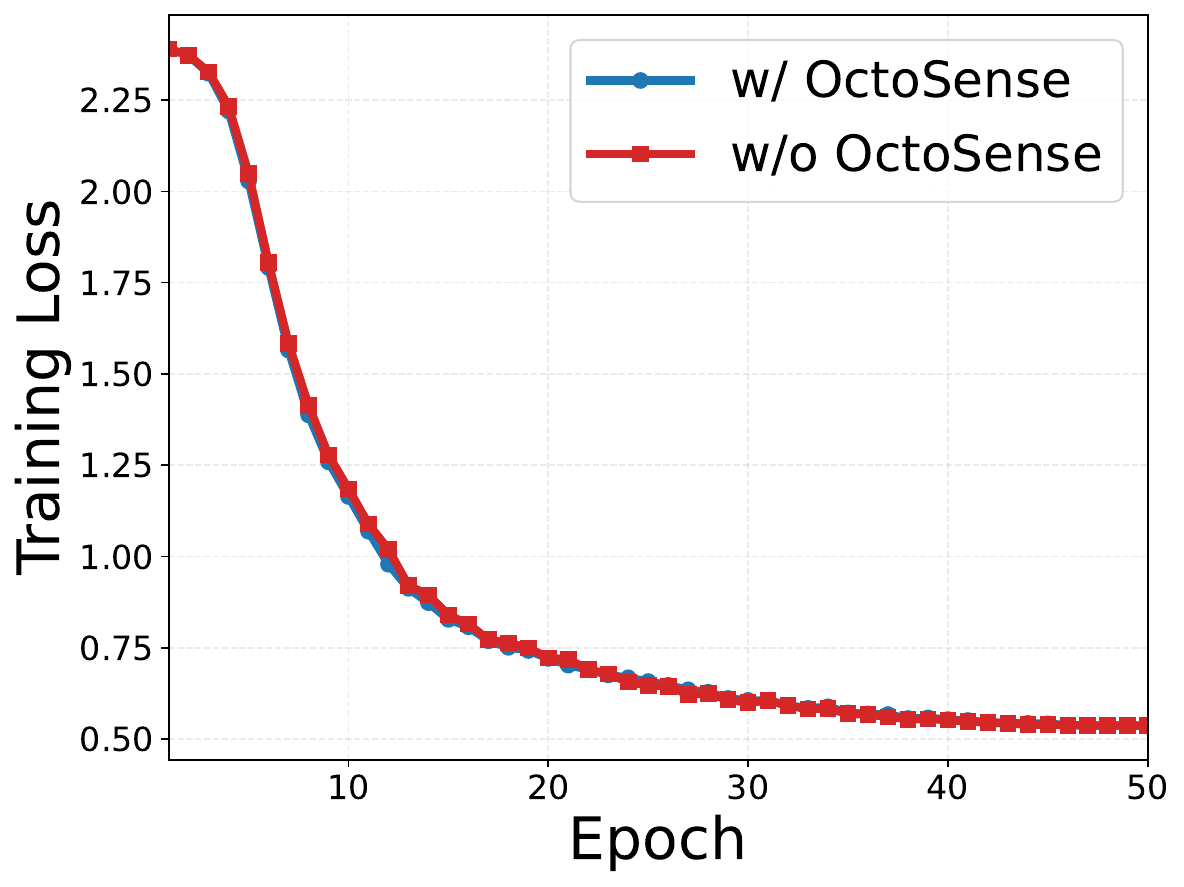}
    }
    \hfill
    \subfloat[RFNet $\times$ WiFi Gait\label{fig:gait_rfnet_loss}]{
        \includegraphics[width=0.22\textwidth]{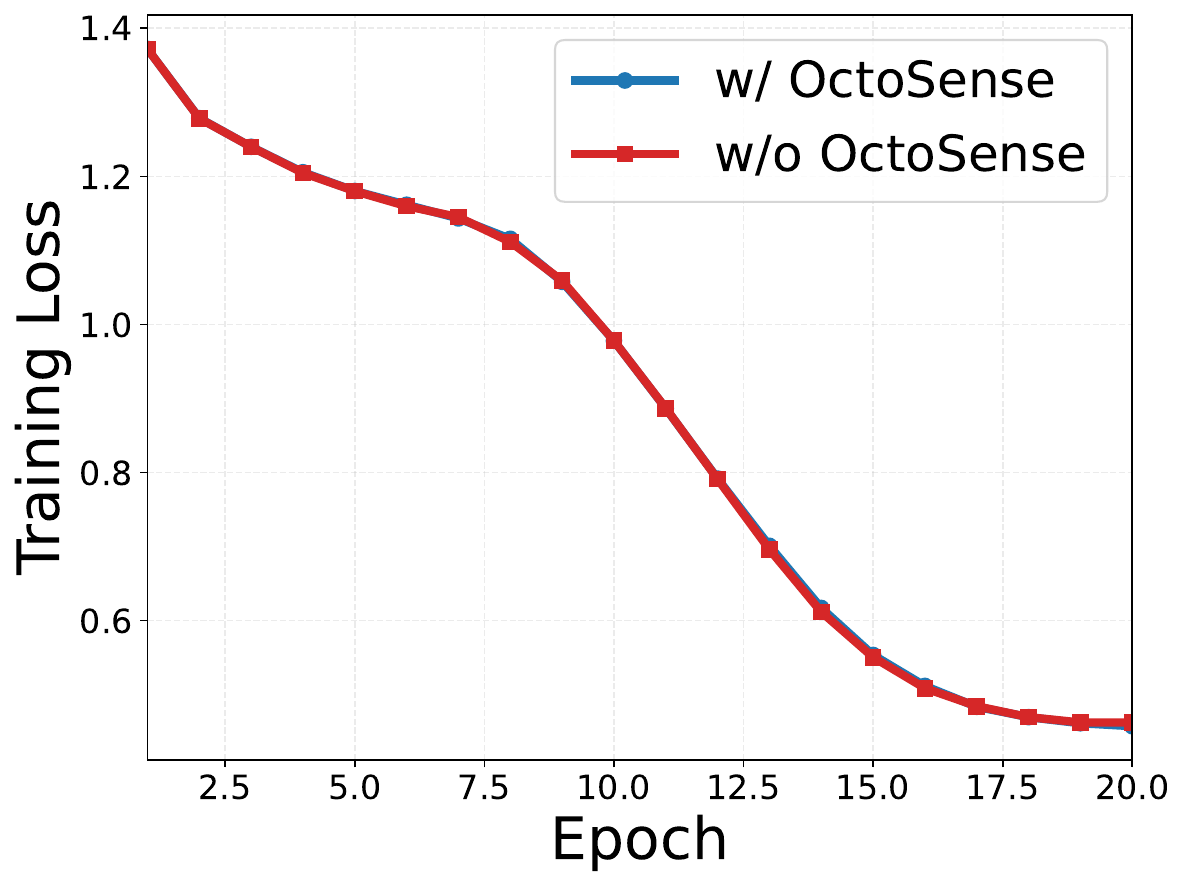}
    }
    \vspace{-0.8\baselineskip}
    \caption{Training loss curves w/ and w/o \sysname{}.}
    \label{fig:training_loss}
    \vspace{-0.1in}
\end{figure}

\begin{table*}[t]
\centering
\scriptsize
\setlength{\tabcolsep}{7.5pt}
\renewcommand{\arraystretch}{1.0}
\caption{Overall performance with and without \sysname.}
\vspace{-0.1in}
\label{tab:overall_comparison}
\begin{tabular}{lllcccccccccccc}
\toprule
\multirow{3}{*}{Dataset} & \multirow{3}{*}{Task} & \multirow{3}{*}{Model}
& \multicolumn{3}{c}{1st-batch Latency (s)}
& \multicolumn{3}{c}{Mean epoch time (s)}
& \multicolumn{3}{c}{Peak Memory (GB)}
& \multicolumn{3}{c}{LLOC$^{*}$} \\
\cmidrule(lr){4-6}
\cmidrule(lr){7-9}
\cmidrule(lr){10-12}
\cmidrule(lr){13-15}
& &
& w/o & w/ & $\Delta$
& w/o & w/ & $\Delta$
& w/o & w/ & $\Delta$
& w/o & w/ & $\Delta$ \\
\midrule

Widar3.0$^{\bigstar}$ & Gesture & STFNet & 4.49 & 3.54 & -0.95 & 182.85 & 172.87 & -9.98 & 3.30 & 3.30 & 0.00 & 159 & 16 & \shortstack{-143\ (-89.94\%)} \\
Widar3.0$^{\bigstar}$ & Gesture & RFNet  & 4.29 & 4.34 & +0.05 & 290.14 & 294.08 & +3.94 & 16.22 & 16.22 & 0.00 & 122 & 15 & \shortstack{-107\ (-87.70\%)} \\
Widar3.0$^{\bigstar}$ & Gesture & UniTS  & 7.00 & 5.62 & -1.38 & 1049.26 & 1046.45 & -2.81 & 7.16 & 7.16 & 0.00 & 150 & 15 & \shortstack{-135\ (-90.00\%)} \\

WiFi Gait$^{\bigstar}$ & Gait & STFNet & 3.06 & 2.83 & -0.23 & 227.87 & 223.54 & -4.33 & 3.33 & 3.33 & 0.00 & 124 & 13 & \shortstack{-111\ (-89.52\%)} \\
WiFi Gait$^{\bigstar}$ & Gait & RFNet  & 2.64 & 2.46 & -0.18 & 361.30 & 361.56 & +0.26 & 19.64 & 19.64 & 0.00 & 87 & 12 & \shortstack{-75\ (-86.21\%)} \\
WiFi Gait$^{\bigstar}$ & Gait & UniTS  & 4.86 & 5.57 & +0.71 & 1241.17 & 1236.96 & -4.21 & 8.35 & 8.35 & 0.00 & 115 & 12 & \shortstack{-103\ (-89.57\%)} \\

MM-Fi$^{\bigstar}$ & HAR & ResNet18 & 0.33 & 0.32 & -0.02 & 191.28 & 247.02 & +55.74 & 0.30 & 0.30 & 0.00 & 118 & 11 & \shortstack{-107\ (-90.68\%)} \\

XRF55$^{\bigstar}$ & HAR & ResNet18 & 0.31 & 0.30 & -0.01 & 51.95 & 61.57 & +9.62 & 0.47 & 0.47 & 0.00 & 114 & 20 & \shortstack{-94\ (-82.46\%)} \\

OctoNet$^{\blacksquare}$ & HAR & ResNet18 & 0.32 & 0.32 & 0.00 & 17.52 & 17.68 & +0.16 & 0.45 & 0.45 & 0.00 & 96 & 10 & \shortstack{-86\ (-89.58\%)} \\

HuPR$^{\blacksquare}$ & Pose & ResNet18 & 0.61 & 1.00 & +0.39 & 46.44 & 58.82 & +12.38 & 0.23 & 0.23 & 0.00 & 122 & 12 & \shortstack{-110\ (-90.16\%)} \\
HuPR$^{\blacksquare}$ & Pose & STFNet & 1.03 & 0.65 & -0.38 & 49.57 & 37.00 & -12.57 & 0.12 & 0.12 & 0.00 & 115 & 23 & \shortstack{-92\ (-80.00\%)} \\
HuPR$^{\blacksquare}$ & Pose & RFNet & 1.15 & 0.80 & -0.35 & 123.91 & 121.52 & -2.39 & 0.47 & 0.47 & 0.00 & 115 & 23 & \shortstack{-92\ (-80.00\%)} \\
HuPR$^{\blacksquare}$ & Pose & UniTS & 1.13 & 1.03 & -0.10 & 146.06 & 145.06 & -1.00 & 2.38 & 2.38 & 0.00 & 115 & 23 & \shortstack{-92\ (-80.00\%)} \\

\bottomrule
\end{tabular}
\vspace{2pt}
\begin{minipage}{0.98\linewidth}
\footnotesize
$^{\bigstar}$ denotes Wi-Fi CSI $^{\blacksquare}$ denotes mmWave.
$\Delta$ denotes differences between w/ and w/o \sysname{}.
$^{*}$Logical Line of Code (LLOC) counts only function calls.
\vspace{-0.2in}
\end{minipage}
\end{table*}

\section{Evaluation}
\label{sec:evaluation}

We evaluate \sysname as an efficient and practical infrastructure for wireless sensing pipelines. Following the design goal of isolating infrastructure effects from model design, we compare paired implementations of the same task: A conventional hand-engineered baseline and an \sysname{}-based version. Across all comparisons, the dataset split, model architecture, training schedule, and evaluation protocol are kept fixed, and only the implementation of data binding and preprocessing differs. 
Our evaluation addresses three questions: \one whether \sysname{} preserves task correctness; \two its end-to-end runtime overhead; and \three how much engineering effort it saves when components are switched or experiments are transferred.

\subsection{\sysname Effectiveness}
\head{Correctness}
\rev{
We first evaluate whether \sysname{} preserves the execution behavior of existing implementations while reducing the amount of user-side pipeline code. We use a paired setting throughout this evaluation: for each dataset and model, we compare the implementation w/o \sysname{} with its counterpart w/ \sysname{} under identical dataset splits, model architectures, training schedules, and evaluation protocols. As summarized in Table~\ref{tab:overall_comparison}, the comparison covers 13 paired configurations across Wi-Fi CSI and mmWave sensing, including gesture recognition on Widar3.0, gait recognition on WiFi Gait, human activity recognition (HAR) on MM-Fi, XRF55, and OctoNet, and pose estimation on HuPR. These tasks are paired with four representative models: ResNet18, STFNet, RFNet, and UniTS.
The results show that using \sysname{} introduces no additional peak memory usage across all configurations, while first-batch latency and epoch time remain dataset- and model-dependent. In several cases, the implementation w/ \sysname{} is slightly faster, such as Widar3.0 with STFNet and HuPR with STFNet, while some ResNet18 configurations incur additional epoch time due to framework-level data orchestration. More importantly, \sysname{} consistently reduces the visible pipeline code: LLOC decreases by 80\% to 91\% across all evaluated configurations. \fig~\ref{fig:training_loss} further shows the training loss curves of two representative paired configurations on WiFi Gait using STFNet and RFNet. In both cases, the loss trajectories remain closely aligned and the test accuracy gap between each paired configuration is at most \revold{0.8\% points}, indicating that \sysname{} preserves the numerical inputs and optimization behavior of the original pipeline.
}

\head{Runtime Overhead}
We next evaluate the runtime cost introduced by \sysname{}. We compare first-batch latency, mean epoch time, and peak memory between the paired implementations. As summarized in \tab~\ref{tab:overall_comparison}, \sysname{} introduces only modest runtime overhead across most task-model pairs. First-batch latency and mean epoch time vary across dataset--model pairs, while the reported peak memory values remain essentially unchanged. These results indicate that the semantic abstraction layer preserves practical execution efficiency.

\head{Reproducibility and Reduced Development Effort}
\rev{
We evaluate whether \sysname{} simplifies wireless sensing benchmark reproduction through an autonomous AI coding case study. Given only a natural-language specification of a representative Deep Wi-Fi sensing pipeline, including Widar3 data loading, CSI preprocessing, spectrogram generation, and ResNet18 forward-pass validation, the same GPT-5.3-codex agent is asked to reproduce the benchmark w/ and w/o \sysname{}.
Without \sysname{}, the agent exhausted all 15 test attempts, consumed 4.95M tokens, and failed to reproduce the benchmark. The failure was traced to an incorrect spectrogram implementation that discarded one side of the STFT spectrum for complex CSI, causing downstream model-input mismatches. With \sysname{}, the agent reproduced the benchmark on its first test attempt using 1.94M tokens, with outputs matching the reference implementation.
These results show that \sysname{} reduces autonomous reproduction effort by replacing benchmark-specific RF preprocessing code with reusable, validated pipeline abstractions. This helps the agent avoid implementation-level RF mistakes that are difficult to diagnose from natural-language specifications alone.
}

\begin{figure}[t]
\centering
    \includegraphics[width=0.45\textwidth]{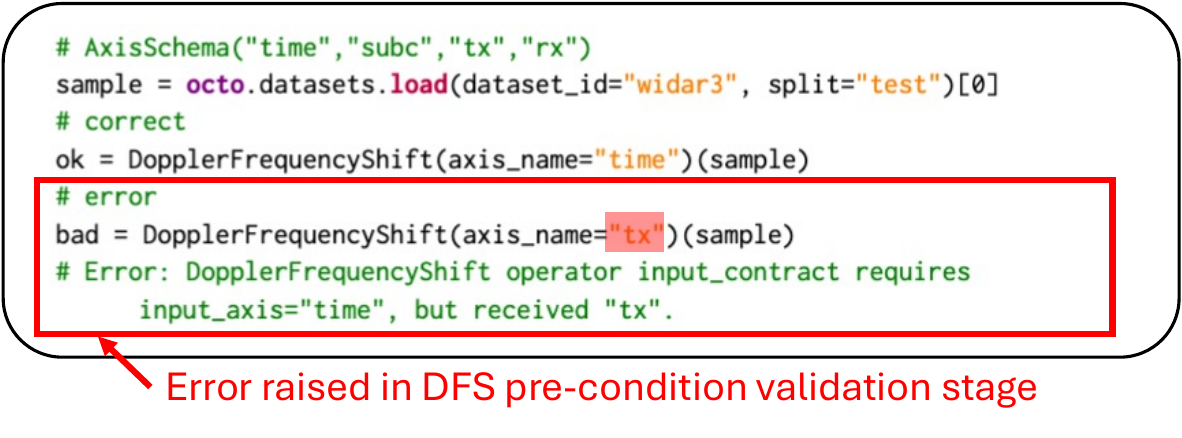}
    \vspace{-1.5\baselineskip}
\caption{Example for operator-axis misuse.}
\label{fig:err_inject}
\vspace{-0.2in}
\end{figure}

\head{Semantic Error Detection}
\rev{
We evaluate whether \sysname{} can detect semantic failures that commonly arise in RF sensing pipelines while leaving the code executable.
We inject eight representative error types into three datasets, Widar3, HuPR, and WiFi Gait: axis ambiguity, signal-domain mismatch, data representation drift, missing operator inputs, missing metadata, split leakage, operator-axis misuse, and model-boundary mismatch. \fig~\ref{fig:err_inject} provides an illustrative example of the injection process, showing how an error can silently pass through ad-hoc code but be exposed once RF semantics and validation checks are made explicit. Across all injected cases, \sysname{} identified every error before model training through schema validation, operator constraints, metadata consistency checks, split validation, and model-boundary verification. These results show that encoding RF semantics makes hidden pipeline assumptions observable and prevents erroneous data from entering downstream training and evaluation.
}

\subsection{Case Study}
\label{sec:case_study}

To demonstrate how \sysname practically solves the fragmented landscape and reproducibility issues in deep wireless sensing, we divide its typical workflow into three progressive stages. These stages show the core capabilities of the framework as a ready-to-use tool, a modular research platform, and an open ecosystem infrastructure. 

\head{Stage 1: Out-of-the-box Baseline Reproduction}
In mature ecosystems like CV, researchers can download and run state-of-the-art pre-trained models with a single click. However, the wireless sensing community lacks this convenience. \sysname provides a simple interface, similar to Hugging Face, as shown in Listing~\ref{lst:case-inference}, allowing beginners or cross-domain researchers to run end-to-end inference with minimal configuration.
By packaging tedious data loading, preprocessing operators, and model weights into a unified pipeline, \sysname removes the parameter configuration barriers when reproducing baseline models.

\begin{pylisting}[label={lst:case-inference}]{Inference pipeline with minimal configuration.}
import octosense as octo
# 1. Load the pre-packaged pipeline and pre-trained STFNet model
pipeline = octo.pipeline.load(
    task="widar3/gesture_classification",
    model="stfnet",
    pretrained=True
)
# 2. Inspect physical semantics of inputs and outputs
pipeline.describe()
# Task: widar3/gesture_classification
# Model: OctoSense-models/stfnet
# Input semantics:
#   Accepts: RadioTensor | WiFiCSIReader
#   Expected axes: ['time', 'subc', 'tx', 'rx']
#   Expected dtype: complex64
# Transforms: Steps: ['Normalize(axis="subc")', ...]
# Model:
#   Layout Type: BLC
#   Axes: ['length', 'channel']
#   Sizes: {'length': 2560, 'channel': 180}
#   Labels: ['Slide', 'Push&Pull', ...]
# Output semantics: Returns: list[dict(index, score)]
# 3. Run instant inference
predictions = pipeline(csi_data)
print(predictions)
# [{'index': 1, 'score': 0.43168},
#  {'index': 5, 'score': 0.21046}, ...]
\end{pylisting}

\head{Stage 2: Modular Benchmarking}
Algorithm researchers often focus on a specific component in the pipeline, such as proposing a new neural network architecture or designing a new signal denoising method. Through a highly decoupled modular design, \sysname allows researchers to freely replace specific components while maintaining the fairness of the benchmark\revold{, with two usage examples below}:

\head{\one Validating Custom Models} 
If researchers design a new sequence model, they can reuse existing standard datasets and processing pipelines, as shown in \revold{Listing~\ref{lst:case-model}}. By using the \texttt{auto\_align} operator, the layout of transformed RF data will be reshaped to the required input dimensions according to the given \texttt{custom\_input\_shape} of the custom model. 

\begin{pylisting}[label={lst:case-model}]{Injecting custom models into standard benchmarks.}
# 1. Load a standard pipeline but with a custom DL model
pipeline = octo.pipeline.load(
    task="widar3/gesture_classification", 
    model=my_custom_transformer, 
    pretrained=False)
# 2. Map RF semantics to the custom model's tensor shape
# e.g. {"channel": "subc", ...}
pipeline.auto_align(custom_input_shape)
# 3. Train the custom model on the standard benchmark
octo.benchmark.train(pipeline)
\end{pylisting}

\head{\two Rebuilding Declarative Pipelines} 
When researchers need to compare different signal processing methods, \sysname provides a clear semantic interface for operator composition. \revold{Listing~\ref{lst:stage2b}} shows a complete benchmarking script for the Widar3 dataset. This declarative design greatly reduces the maintenance cost of comparative experiments. Researchers no longer need to search for implicit tensor dimensions in complex script files. They only need to modify the \texttt{compose} list to complete complex ablation studies.

\begin{pylisting}[label={lst:stage2b}]{Declarative signal processing pipeline.}
# 1. Load RF datasets (auto-downloading if needed).
datasets = octo.dataset.load("widar3", split="train+val")
# 2. Define signal processing pipeline
transforms = octo.transform.compose([
    octo.transform.csi.pilot_interp(method="linear"),
    octo.transform.axis_norm(axis="subc", method="l2"), 
    octo.transform.rx_conjugate(metric="top-variance"),
    octo.transform.PCA(axis="subc", n=1),
    octo.transform.stft(axis="time", n_fft=128, hop=64),
    octo.transform.magnitude(),
])
# 3. Build dataloaders and run training
dataloaders = octo.loader.build(datasets, transforms)
model = octo.models.image.resnet18()
octo.benchmark.train(model, dataloaders)
\end{pylisting}

\begin{pylisting}[label={lst:case-dataset}]{Integrating custom dataset into the ecosystem.}
# 1. Define custom reader to load data from new hardware
# class MyCustomReader(octo.reader.BaseCSIReader): ...
reader = MyCustomReader(target_length=2560)
# 2. Custom reader converts raw data into shared-semantic tensors
csi_tensor = reader.read("./lab/*.dat")
# dims = ["time", "subc", "tx", "rx"],
# metadata = {"center_freq": "5.8GHz", ...}
# 3. Wrap semantic RF tensors into the dataset abstraction
labels = np.load("./lab/labels.npy")
dataset = octo.dataset.from_tensor(csi_tensor, labels)
# 4. Reuse existing standard operators and models
pipeline = octo.pipeline.load(
    task="widar3_default",
    dataset=dataset,
    model="stfnet"
)
pred = octo.benchmark.evaluate(pipeline)

\end{pylisting}

\head{Stage 3: Ecosystem Integration}
Real-world wireless sensing research constantly generates data from new hardware platforms and private datasets. \sysname acts as a universal infrastructure rather than a closed tool. 
As shown in \revold{Listing~\ref{lst:case-dataset}}, users can define custom readers through a common interface, map raw data into shared RF semantics, and wrap them into a standardized dataset abstraction. Once integrated, these data can directly reuse existing operators, models, and benchmark pipelines.

By enforcing explicit physical semantics at data ingestion, \sysname transforms isolated laboratory data into interoperable ecosystem modules. This connects the full chain from hardware collection to deep learning models, laying a standard foundation for sharing and reusing heterogeneous datasets in the future.

\section{Discussions and Future Work}
\label{sec:discussion}

\head{Scope and Applicability}
OctoSense primarily serves DWS workflows, where data scale and model iteration are the main challenges. For conventional signal-processing-centric sensing, our contribution focuses on the Data and Transform layers. By providing a unified metadata specification, we centralize one-time reader adaptation and reduce repeated manual parsing in downstream workflows. 
Our work is \textit{orthogonal} to specific algorithm innovations. OctoSense does not directly solve open research problems, such as model generalization or domain adaptation, but instead provides a standardized infrastructure to facilitate the development of these solutions. By ensuring different models run on identical pipelines, researchers can conduct fair benchmarks and focus on their core algorithmic contributions.

\head{Roadmap and Community Governance}
We will continue to expand OctoSense in several key directions. First, we plan to introduce automatic metadata inference to reduce manual configuration. Second, we aim to support streaming and online pipelines, extending the framework from offline training to real-time deployment. On the ecosystem side, we will develop versioning and migration tools to maintain compatibility between existing datasets and future framework updates. OctoSense has already been open-sourced, and we will further support its growth through a clear governance mechanism that encourages community contributions of new operators, datasets, and pre-trained model families.

\section{Related Work}
\label{sec:related-work}

\head{Ecosystems in Adjacent Domains}
CV and NLP have achieved sustained prosperity over the past decade. This success is not merely due to individual powerful models. It stems from a shared public foundation, which includes widely available open datasets, unified data management, mature development frameworks, and standardized benchmark systems. To integrate domain knowledge into deep learning frameworks, various fields have developed specialized ecosystems. These include computer vision~\cite{torchvision}, audio and speech processing~\cite{torchaudio, lhotse}, geospatial research~\cite{torchgeo}, medical imaging~\cite{cardoso2022monai}, recommender systems~\cite{torchrec}, physics simulations~\cite{modulus}, biology research~\cite{torchdrug}, and graph processing~\cite{pyg, dgl}. 
These ecosystems share a common pattern: they unify data loading, domain-specific preprocessing, and model integration under a single API. Inspired by them, \sysname{} addresses the unique challenges of RF data by introducing unified RF semantics and reusable pipelines for dataset switching, model development, and benchmarking.

\head{RF Sensing Ecosystem}
Despite the rapid growth of wireless sensing in models, algorithms, and applications, the research community remains highly fragmented across different stages of the pipeline. At the hardware extraction level, tools like PicoScenes~\cite{picoscenes} and CSIKit~\cite{csikit} support data parsing and processing across multiple hardware formats. However, they operate purely at the parsing level. They produce basic numerical arrays without physical metadata or semantic axis names, meaning a user who accidentally transposes dimensions will receive no warning. Similarly, for radar signals, OpenRadar~\cite{openradar2019} provides signal processing wrappers, but it relies on traditional numerical array operations. It lacks physical metadata tracking and semantic axis validation, making complex deep learning pipelines prone to silent alignment errors. Moving up to the dataset and application level, an increasing number of public resources, such as Widar3~\cite{widar3}, MM-Fi~\cite{mmfi}, XRF55~\cite{xrf55}, and OctoNet~\cite{octonet}, have become available. Researchers continually propose task-specific models for these datasets. However, these models are almost always bundled with heavily customized, specific preprocessing pipelines. Finally, for benchmarking, early attempts like SenseFi~\cite{sensefi} and CSI-Bench~\cite{csibench} aim to establish unified baselines. While SenseFi standardizes data loading and model training for WiFi CSI, it operates primarily at the model level. It unifies tensor shapes but does not attach semantic meaning to tensor axes or enforce preprocessing contracts. Unlike image dimensions, which are standard and rarely confused, RF tensor dimensions, such as time, subcarrier, antenna, and chirp, vary wildly across hardware and are frequently misaligned. Consequently, datasets, models, and benchmarks continue to evolve in isolated, local loops without a unifying common infrastructure.

\head{Our Positioning and Contribution}
To directly address this fragmentation, \sysname{} provides a comprehensive open-source framework that seamlessly connects dataset support, model development, and baseline benchmarking. We establish a common core capable of sustaining data, models, and workflows simultaneously. By introducing a unified data language, \sysname standardizes diverse hardware formats into reliable data structures. Furthermore, by designing reusable operator interfaces, we simplify complex signal processing steps into composable pipelines. To the best of our knowledge, \sysname{} is the first work to explicitly integrate unified data representations, semantic operators, and reproducible workflows into a single platform. By providing this software-side semantic safety layer, \sysname{} allows future research to accumulate around a shared system boundary, empowering researchers to focus on algorithmic innovation rather than repetitive reimplementations.

\section{Conclusion}
\label{sec:conclusion}
Wireless sensing holds significant promise as a non-intrusive modality for enabling physical intelligence. 
However, its progress has been hindered by fragmented datasets, incompatible preprocessing pipelines, and the absence of standardized benchmarking practices, collectively creating a substantial barrier to reproducible and cumulative research. 
This paper presents \sysname, a holistic infrastructure to overcome these challenges through three key components: a unified data abstraction, standardized signal operators, and a specification-driven benchmarking engine. 
With \sysname, we aim to shift the wireless sensing research landscape into a unified and sustainable ecosystem where community efforts are interoperable, reusable, and reproducible, and we hope it will serve as a foundation for continued evolution.

\bibliographystyle{ACM-Reference-Format}
\bibliography{refs}

@misc{liangVisionMultisensoryIntelligence2026,
  title = {A {{Vision}} for {{Multisensory Intelligence}}: {{Sensing}}, {{Science}}, and {{Synergy}}},
  shorttitle = {A {{Vision}} for {{Multisensory Intelligence}}},
  author = {Liang, Paul Pu},
  year = 2026,
  month = jan,
  primaryclass = {cs.LG},
  publisher = {arXiv},
  doi = {10.48550/arXiv.2601.04563},
  urldate = {2026-01-22}
}

@article{guarino2026survey,
  title={A Survey on {CSI}-Based {Wi-Fi} Sensing Datasets and Models with a Focus on Reproducibility},
  author={Guarino, Idio and Carra, Damiano and Cominelli, Marco and Gringoli, Francesco and Cigno, Renato Lo},
  journal={Computer Communications},
  pages={108431},
  year={2026},
  publisher={Elsevier}
}

@article{varga2024mitigating,
  title={Mitigating Data Leakage in a {WiFi} {CSI} Benchmark for Human Action Recognition},
  author={Varga, Domonkos},
  journal={Sensors},
  volume={24},
  number={24},
  pages={8201},
  year={2024},
  publisher={MDPI}
}

@article{Hou2024RFBoostUA,
  title={{RFBoost}: Understanding and Boosting Deep {WiFi} Sensing via Physical Data Augmentation},
  author={Weiying Hou and Chenshu Wu},
  journal={Proceedings of the ACM on Interactive, Mobile, Wearable and Ubiquitous Technologies},
  year={2024},
  volume={8},
  pages={1 - 26},
  url={https://api.semanticscholar.org/CorpusID:269864274}
}

@article{Kapoor2023LeakageAT,
  title={Leakage and the reproducibility crisis in machine-learning-based science},
  author={Sayash Kapoor and Arvind Narayanan},
  journal={Patterns},
  year={2023},
  volume={4},
  url={https://api.semanticscholar.org/CorpusID:260663045}
}

@ARTICLE{widar3,
  author={Zhang, Yi and Zheng, Yue and Qian, Kun and Zhang, Guidong and Liu, Yunhao and Wu, Chenshu and Yang, Zheng},
  journal={IEEE Transactions on Pattern Analysis and Machine Intelligence}, 
  title={Widar3.0: Zero-Effort Cross-Domain Gesture Recognition With Wi-Fi}, 
  year={2022},
  volume={44},
  number={11},
  pages={8671-8688},
  doi={10.1109/TPAMI.2021.3105387}}

@misc{Wightman2021ResNetSB,
  title={ResNet Strikes Back: An Improved Training Procedure in timm},
  author={Ross Wightman and Hugo Touvron and Herv{\'e} J{\'e}gou},
  year={2021},
  primaryclass={cs.CV},
  publisher={arXiv},
  doi={10.48550/arXiv.2110.00476}
}

@ARTICLE{deepsense,
  author={Alkhateeb, Ahmed and Charan, Gouranga and Osman, Tawfik and Hredzak, Andrew and Morais, Joao and Demirhan, Umut and Srinivas, Nikhil},
  journal={IEEE Communications Magazine}, 
  title={DeepSense 6G: A Large-Scale Real-World Multi-Modal Sensing and Communication Dataset}, 
  year={2023},
  volume={61},
  number={9},
  pages={122-128},
  doi={10.1109/MCOM.006.2200730}}

@InProceedings{hupr,
    author    = {Lee, Shih-Po and Kini, Niraj Prakash and Peng, Wen-Hsiao and Ma, Ching-Wen and Hwang, Jenq-Neng},
    title     = {HuPR: A Benchmark for Human Pose Estimation Using Millimeter Wave Radar},
    booktitle = {Proceedings of the IEEE/CVF Winter Conference on Applications of Computer Vision (WACV)},
    month     = {January},
    year      = {2023},
    pages     = {5715--5724},
    publisher = {IEEE},
    location  = {Waikoloa, HI, USA},
    url       = {https://openaccess.thecvf.com/content/WACV2023/html/Lee_HuPR_A_Benchmark_for_Human_Pose_Estimation_Using_Millimeter_Wave_WACV_2023_paper.html}
}

@inproceedings{cuhkx,
  title = {A Large-Scale Multimodal Dataset and Benchmarks for Human Activity Scene Understanding and Reasoning},
  author = {Siyang Jiang and Mu Yuan and Xiang Ji and Bufang Yang and Zeyu Liu and Lilin Xu and Yang Li and Yuting He and Liran Dong and Wenrui Lu and Zhenyu Yan and Xiaofan Jiang and Wei Gao and Hongkai Chen and Guoliang Xing},
  year = {2026},
  url = {https://doi.org/10.1145/3745756.3809209},
  booktitle = {Proceedings of the 24th Annual International Conference on Mobile Systems, Applications and Services},
  series = {MobiSys '26},
  pages = {352--370},
  numpages = {19},
  publisher = {Association for Computing Machinery},
  address = {New York, NY, USA},
  location = {Cambridge, United Kingdom},
  isbn = {9798400720277},
  doi = {10.1145/3745756.3809209}
}

@article{xrf_v2,
author = {Lan, Bo and Li, Pei and Yin, Jiaxi and Song, Yunpeng and Wang, Ge and Ding, Han and Han, Jinsong and Wang, Fei},
title = {XRF V2: A Dataset for Action Summarization with Wi-Fi Signals, and IMUs in Phones, Watches, Earbuds, and Glasses},
year = {2025},
issue_date = {September 2025},
publisher = {Association for Computing Machinery},
address = {New York, NY, USA},
volume = {9},
number = {3},
url = {https://doi.org/10.1145/3749521},
doi = {10.1145/3749521},
journal = {Proceedings of the ACM on Interactive, Mobile, Wearable and Ubiquitous Technologies},
month = sep,
articleno = {98},
numpages = {41}
}

@inproceedings{csibench,
  title = {{CSI-Bench}: A Large-Scale In-the-Wild Dataset for Multi-task {WiFi} Sensing},
  author = {Guozhen Zhu and Yuqian Hu and Weihang Gao and Wei-Hsiang Wang and Beibei Wang and K. J. Ray Liu},
  year = {2025},
  url = {https://proceedings.neurips.cc/paper_files/paper/2025/hash/f7c68372b3d39e8d7a093eb2edaaad87-Abstract-Datasets_and_Benchmarks_Track.html},
  booktitle = {Advances in Neural Information Processing Systems 38},
  pages = {187882--187907},
  publisher = {Neural Information Processing Systems Foundation, Inc.},
  location = {San Diego, CA, USA and Mexico City, Mexico},
  doi = {10.52202/085713-5648}
}

@inproceedings{mmfi,
  title = {MM-Fi: Multi-Modal Non-Intrusive 4D Human Dataset for Versatile Wireless Sensing},
  author = {Jianfei Yang and He Huang and Yunjiao Zhou and Xinyan Chen and Yuecong Xu and Shenghai Yuan and Han Zou and Chris Xiaoxuan Lu and Lihua Xie},
  year = {2023},
  url = {https://proceedings.neurips.cc/paper_files/paper/2023/hash/3baf7a39d07e9f4f1e258a412df94521-Abstract-Datasets_and_Benchmarks.html},
  booktitle = {Advances in Neural Information Processing Systems 36 (NeurIPS 2023)},
  pages = {18756--18768},
  publisher = {Neural Information Processing Systems Foundation, Inc.},
  location = {New Orleans, LA, USA},
  doi = {10.52202/075280-0822}
}

@article{sensefi,
  title = {{SenseFi}: A Library and Benchmark on Deep-Learning-Empowered {WiFi} Human Sensing},
  author = {Jianfei Yang and Xinyan Chen and Han Zou and Chris Xiaoxuan Lu and Dazhuo Wang and Sumei Sun and Lihua Xie},
  journal = {Patterns},
  year = {2023},
  volume = {4},
  url = {https://doi.org/10.1016/j.patter.2023.100703},
  number = {3},
  pages = {100703},
  doi = {10.1016/j.patter.2023.100703}
}

@article{picoscenes,
  title = {Eliminating the Barriers: Demystifying Wi-Fi Baseband Design and Introducing the PicoScenes Wi-Fi Sensing Platform},
  author = {Zhiping Jiang and Tom H. Luan and Xincheng Ren and Dongtao Lv and Han Hao and Jing Wang and Kun Zhao and Wei Xi and Yueshen Xu and Rui Li},
  journal = {IEEE Internet of Things Journal},
  year = {2022},
  volume = {9},
  url = {https://doi.org/10.1109/JIOT.2021.3104666},
  number = {6},
  pages = {4476--4496},
  doi = {10.1109/JIOT.2021.3104666}
}

@misc{openradar2019,
  author = {Pan, Edwin and Tang, Jingning and Kosaka, Dash and Yao, Ruihao and Gupta, Arjun},
  title = {OpenRadar},
  year = {2019},
  publisher = {GitHub},
  journal = {GitHub repository},
  howpublished = {\url{https://github.com/presenseradar/openradar}}
}

@article{xrf55,
author = {Wang, Fei and Lv, Yizhe and Zhu, Mengdie and Ding, Han and Han, Jinsong},
title = {XRF55: A Radio Frequency Dataset for Human Indoor Action Analysis},
year = {2024},
issue_date = {March 2024},
publisher = {Association for Computing Machinery},
address = {New York, NY, USA},
volume = {8},
number = {1},
url = {https://doi.org/10.1145/3643543},
doi = {10.1145/3643543},
journal = {Proceedings of the ACM on Interactive, Mobile, Wearable and Ubiquitous Technologies},
month = mar,
articleno = {21},
numpages = {34}
}

@inproceedings{octonet,
  title     = {OctoNet: A Large-Scale Multi-Modal Dataset for Human Activity Understanding Grounded in Motion-Captured 3D Pose Labels},
  author    = {Yuan, Dongsheng and Zhang, Xie and Hou, Weiying and Lyu, Sheng and Yu, Yuemin and Yu, Luca Jiang-Tao and Li, Chengxiao and Wu, Chenshu},
  booktitle = {Advances in Neural Information Processing Systems 38},
  pages     = {15609--15633},
  publisher = {Neural Information Processing Systems Foundation, Inc.},
  doi       = {10.52202/085713-0469},
  year      = {2025},
  location  = {San Diego, CA, USA and Mexico City, Mexico},
  note      = {NeurIPS 2025 Datasets and Benchmarks Track},
  url       = {https://proceedings.neurips.cc/paper_files/paper/2025/hash/14950adea25005f89545688fe97fe5ea-Abstract-Datasets_and_Benchmarks_Track.html}
}

@misc{sdp8,
  author       = {{SDP Competition Platform}},
  title        = {{SSC2025 Competition - Software Defined Network Contest}},
  howpublished = {\url{https://sdp8.net/}},
  note         = {Accessed: 2026-03-14},
  year = {2025}
}

@misc{xarray,
  author       = {{Xarray}},
  title        = {{xarray.DataArray}},
  howpublished = {\url{https://docs.xarray.dev/en/stable/generated/xarray.DataArray.html}},
  note         = {Accessed: 2026-03-14},
  year = {2026}
}

@misc{ieeedataport,
  author       = {{IEEE DataPort}},
  title        = {{IEEE DataPort}},
  howpublished = {\url{https://ieee-dataport.org/}},
  note         = {Accessed: 2026-03-14},
  year = {2026}
}

@misc{cardoso2022monai,
  title={{MONAI}: An Open-Source Framework for Deep Learning in Healthcare},
  author={Cardoso, M Jorge and Li, Wenqi and Brown, Richard and Ma, Nic and Kerfoot, Eric and Wang, Yiheng and Murrey, Benjamin and Myronenko, Andriy and Zhao, Can and Yang, Dong and others},
  year={2022},
  primaryclass={cs.LG},
  publisher={arXiv},
  doi={10.48550/arXiv.2211.02701}
}

@inproceedings{UniTS,
author = {Li, Shuheng and Chowdhury, Ranak Roy and Shang, Jingbo and Gupta, Rajesh K. and Hong, Dezhi},
title = {UniTS: Short-Time Fourier Inspired Neural Networks for Sensory Time Series Classification},
year = {2021},
isbn = {9781450390972},
publisher = {Association for Computing Machinery},
address = {New York, NY, USA},
url = {https://doi.org/10.1145/3485730.3485942},
doi = {10.1145/3485730.3485942},
booktitle = {Proceedings of the 19th ACM Conference on Embedded Networked Sensor Systems},
series = {SenSys '21},
location = {Coimbra, Portugal},
pages = {234--247}
}

@inproceedings{dingRFnetUnifiedMetalearning2020,
  title = {{{RF-Net}}: A Unified Meta-Learning Framework for {{RF-enabled}} One-Shot Human Activity Recognition},
  shorttitle = {{{RF-Net}}},
  booktitle = {Proceedings of the 18th {{Conference}} on {{Embedded Networked Sensor Systems}}},
  author = {Ding, Shuya and Chen, Zhe and Zheng, Tianyue and Luo, Jun},
  year = {2020},
  month = nov,
  series = {{{SenSys}} '20},
  pages = {517--530},
  publisher = {Association for Computing Machinery},
  address = {New York, NY, USA},
  location = {Virtual Event, Japan},
  doi = {10.1145/3384419.3430735},
  url = {https://doi.org/10.1145/3384419.3430735},
  urldate = {2022-12-04},
  isbn = {978-1-4503-7590-0}
}

@inproceedings{zhao2018rf,
  title={{RF}-based 3D Skeletons},
  author={Zhao, Mingmin and Tian, Yonglong and Zhao, Hang and Alsheikh, Mohammad Abu and Li, Tianhong and Hristov, Rumen and Kabelac, Zachary and Katabi, Dina and Torralba, Antonio},
  booktitle={Proceedings of the 2018 Conference of the ACM Special Interest Group on Data Communication},
  pages={267--281},
  year={2018},
  series={SIGCOMM '18},
  publisher={Association for Computing Machinery},
  address={New York, NY, USA},
  location={Budapest, Hungary},
  doi={10.1145/3230543.3230579},
  url={https://doi.org/10.1145/3230543.3230579}
}

@inproceedings{zhengMoReFiMotionrobustFinegrained2021,
  title = {{{MoRe-Fi}}: {{Motion-robust}} and {{Fine-grained Respiration Monitoring}} via {{Deep-Learning UWB Radar}}},
  shorttitle = {{{MoRe-Fi}}},
  booktitle = {Proceedings of the 19th {{ACM Conference}} on {{Embedded Networked Sensor Systems}}},
  author = {Zheng, Tianyue and Chen, Zhe and Zhang, Shujie and Cai, Chao and Luo, Jun},
  year = {2021},
  month = nov,
  series = {{{SenSys}} '21},
  pages = {111--124},
  publisher = {Association for Computing Machinery},
  address = {New York, NY, USA},
  location = {Coimbra, Portugal},
  doi = {10.1145/3485730.3485932},
  urldate = {2022-12-02},
  isbn = {978-1-4503-9097-2}
}

@inproceedings{yaoSTFNetsLearningSensing2019,
  title = {{{STFNets}}: {{Learning Sensing Signals}} from the {{Time-Frequency Perspective}} with {{Short-Time Fourier Neural Networks}}},
  shorttitle = {{{STFNets}}},
  booktitle = {The {{World Wide Web Conference}}},
  author = {Yao, Shuochao and Piao, Ailing and Jiang, Wenjun and Zhao, Yiran and Shao, Huajie and Liu, Shengzhong and Liu, Dongxin and Li, Jinyang and Wang, Tianshi and Hu, Shaohan and Su, Lu and Han, Jiawei and Abdelzaher, Tarek},
  year = {2019},
  month = may,
  series = {{{WWW}} '19},
  pages = {2192--2202},
  publisher = {Association for Computing Machinery},
  address = {New York, NY, USA},
  location = {San Francisco, CA, USA},
  doi = {10.1145/3308558.3313426},
  urldate = {2022-12-02},
  isbn = {978-1-4503-6674-8}
}

@inproceedings{chi2024rf,
  title={RF-Diffusion: Radio Signal Generation via Time-Frequency Diffusion},
  author={Chi, Guoxuan and Yang, Zheng and Wu, Chenshu and Xu, Jingao and Gao, Yuchong and Liu, Yunhao and Han, Tony Xiao},
  booktitle={Proceedings of the 30th Annual International Conference on Mobile Computing and Networking},
  pages={77--92},
  year={2024},
  series={MobiCom '24},
  publisher={Association for Computing Machinery},
  address={New York, NY, USA},
  location={Washington, DC, USA},
  doi={10.1145/3636534.3649348},
  url={https://doi.org/10.1145/3636534.3649348}
}

@article{zhangWiDetectRobustMotion2019,
  title = {{{WiDetect}}: {{Robust Motion Detection}} with a {{Statistical Electromagnetic Model}}},
  shorttitle = {{{WiDetect}}},
  author = {Zhang, Feng and Wu, Chenshu and Wang, Beibei and Lai, Hung-Quoc and Han, Yi and Liu, K. J. Ray},
  year = 2019,
  month = sep,
  journal = {Proceedings of the ACM on Interactive, Mobile, Wearable and Ubiquitous Technologies},
  volume = {3},
  number = {3},
  pages = {122:1--122:24},
  doi = {10.1145/3351280},
  urldate = {2021-09-21}
}

@article{yang2022artificial,
  title={Artificial intelligence-enabled detection and assessment of Parkinson’s disease using nocturnal breathing signals},
  author={Yang, Yuzhe and Yuan, Yuan and Zhang, Guo and Wang, Hao and Chen, Ying-Cong and Liu, Yingcheng and Tarolli, Christopher G and Crepeau, Daniel and Bukartyk, Jan and Junna, Mithri R and others},
  journal={{Nature Medicine}},
  volume={28},
  number={10},
  pages={2207--2215},
  year={2022},
  publisher={Nature Publishing Group US New York}
}

@inproceedings{lai2024enabling,
  title={Enabling Visual Recognition at Radio Frequency},
  author={Lai, Haowen and Luo, Gaoxiang and Liu, Yifei and Zhao, Mingmin},
  booktitle={Proceedings of the 30th Annual International Conference on Mobile Computing and Networking},
  pages={388--403},
  year={2024},
  series={MobiCom '24},
  publisher={Association for Computing Machinery},
  address={New York, NY, USA},
  location={Washington, DC, USA},
  doi={10.1145/3636534.3649369},
  url={https://doi.org/10.1145/3636534.3649369}
}

@inproceedings{boroushaki2021rfusion,
  title={{RFusion}: Robotic Grasping via {RF-Visual} Sensing and Learning},
  author={Boroushaki, Tara and Perper, Isaac and Nachin, Mergen and Rodriguez, Alberto and Adib, Fadel},
  booktitle={Proceedings of the 19th ACM Conference on Embedded Networked Sensor Systems},
  pages={192--205},
  year={2021},
  series={SenSys '21},
  publisher={Association for Computing Machinery},
  address={New York, NY, USA},
  location={Coimbra, Portugal},
  doi={10.1145/3485730.3485944},
  url={https://doi.org/10.1145/3485730.3485944}
}

@inproceedings{deng2009imagenet,
  title={{ImageNet}: A Large-Scale Hierarchical Image Database},
  author={Deng, Jia and Dong, Wei and Socher, Richard and Li, Li-Jia and Li, Kai and Fei-Fei, Li},
  booktitle={2009 IEEE Conference on Computer Vision and Pattern Recognition (CVPR)},
  pages={248--255},
  year={2009},
  publisher={IEEE},
  location={Miami, FL, USA},
  doi={10.1109/CVPR.2009.5206848},
  url={https://doi.org/10.1109/CVPR.2009.5206848}
}

@electronic{csikit,
      author = {Forbes, Glenn},
      title = {{CSIKit}: Python {CSI} Processing and Visualisation Tools for Commercial Off-the-Shelf Hardware},
      url = {https://github.com/Gi-z/CSIKit},
      year = {2021}
  }

@manual{cufft,
title = {cuFFT Documentation},
author = {{NVIDIA Corporation}},
organization = {NVIDIA Corporation},
year = {2026},
note = {Release 13.2}
}

@article{5300,
  title={Tool release: gathering 802.11n traces with channel state information},
  author={Daniel Halperin and Wenjun Hu and Anmol Sheth and David Wetherall},
  journal={Comput. Commun. Rev.},
  year={2011},
  volume={41},
  pages={53},
  url={https://api.semanticscholar.org/CorpusID:13561174}
}

@inproceedings{resnet,
  title={Deep Residual Learning for Image Recognition},
  author={He, Kaiming and Zhang, Xiangyu and Ren, Shaoqing and Sun, Jian},
  booktitle={2016 IEEE Conference on Computer Vision and Pattern Recognition (CVPR)},
  pages={770--778},
  year={2016},
  publisher={IEEE},
  location={Las Vegas, NV, USA},
  doi={10.1109/CVPR.2016.90},
  url={https://doi.org/10.1109/CVPR.2016.90}
}

@misc{torchaudio,
      title={TorchAudio 2.1: Advancing speech recognition, self-supervised learning, and audio processing components for PyTorch},
      author={Jeff Hwang and Moto Hira and Caroline Chen and Xiaohui Zhang and Zhaoheng Ni and Guangzhi Sun and Pingchuan Ma and Ruizhe Huang and Vineel Pratap and Yuekai Zhang and Anurag Kumar and Chin-Yun Yu and Chuang Zhu and Chunxi Liu and Jacob Kahn and Mirco Ravanelli and Peng Sun and Shinji Watanabe and Yangyang Shi and Yumeng Tao and Robin Scheibler and Samuele Cornell and Sean Kim and Stavros Petridis},
      year={2023},
      primaryclass={eess.AS},
      publisher={arXiv},
      doi={10.48550/arXiv.2310.17864}
}

@inproceedings{pytorch2019,
  title = {{PyTorch}: An Imperative Style, High-Performance Deep Learning Library},
  author = {Paszke, Adam and Gross, Sam and Massa, Francisco and Lerer, Adam and Bradbury, James and Chanan, Gregory and Killeen, Trevor and Lin, Zeming and Gimelshein, Natalia and Antiga, Luca and Desmaison, Alban and Kopf, Andreas and Yang, Edward and DeVito, Zachary and Raison, Martin and Tejani, Alykhan and Chilamkurthy, Sasank and Steiner, Benoit and Fang, Lu and Bai, Junjie and Chintala, Soumith},
  booktitle = {Advances in Neural Information Processing Systems 32 (NeurIPS 2019)},
  editor = {H. Wallach and H. Larochelle and A. Beygelzimer and F. d'Alch{\'e}-Buc and E. Fox and R. Garnett},
  pages = {8024--8035},
  publisher = {Curran Associates, Inc.},
  location = {Vancouver, Canada},
  year = {2019},
  url = {https://proceedings.neurips.cc/paper_files/paper/2019/file/bdbca288fee7f92f2bfa9f7012727740-Paper.pdf}
}

@software{torchvision,
  author = {{TorchVision maintainers and contributors}},
  title = {TorchVision: PyTorch's Computer Vision Library},
  year = {2016},
  url = {https://github.com/pytorch/vision},
  note = {GitHub repository}
}

@inproceedings{lhotse,
  title = {Lhotse: A Speech Data Representation Library for the Modern Deep Learning Ecosystem},
  author = {Zelasko, Piotr and Povey, Daniel and Trmal, Jan and Khudanpur, Sanjeev},
  booktitle = {NeurIPS 2021 Workshop on Data-Centric AI},
  year = {2021},
  location = {Virtual Event},
  primaryclass = {cs.SD},
  doi = {10.48550/arXiv.2110.12561}
}

@misc{torchgeo,
  title = {TorchGeo: Deep Learning with Geospatial Data},
  author = {Stewart, Adam J. and Robinson, Caleb and Corley, Isaac A. and Ortiz, Anthony and Lavista Ferres, Juan M. and Banerjee, Arindam},
  year = {2021},
  primaryclass = {cs.CV},
  publisher = {arXiv},
  doi = {10.48550/arXiv.2111.08872}
}

@inproceedings{torchrec,
  author = {Ivchenko, Dmytro and Van Der Staay, Dennis and Taylor, Colin and Liu, Xing and Feng, Will and Kindi, Rahul and Sudarshan, Anirudh and Sefati, Shahin},
  title = {TorchRec: a PyTorch Domain Library for Recommendation Systems},
  year = {2022},
  isbn = {9781450392785},
  publisher = {Association for Computing Machinery},
  address = {New York, NY, USA},
  url = {https://doi.org/10.1145/3523227.3547387},
  doi = {10.1145/3523227.3547387},
  booktitle = {Proceedings of the 16th ACM Conference on Recommender Systems},
  pages = {482--483},
  location = {Seattle, WA, USA},
  series = {RecSys '22}
}

@software{modulus,
  author = {{NVIDIA Modulus Team}},
  title = {NVIDIA Modulus},
  year = {2022},
  url = {https://docs.nvidia.com/deeplearning/modulus/modulus-v2209/user_guide/basics/modulus_overview.html},
  note = {Open-source deep-learning framework for physics-informed and data-driven machine learning; project now continued as NVIDIA PhysicsNeMo}
}

@misc{torchdrug,
  title = {TorchDrug: A Powerful and Flexible Machine Learning Platform for Drug Discovery},
  author = {Zhu, Zhaocheng and Shi, Chence and Zhang, Zuobai and Liu, Shengchao and Xu, Minghao and Yuan, Xinyu and Zhang, Yangtian and Chen, Junkun and Cai, Huiyu and Lu, Jiarui and Ma, Chang and Liu, Runcheng and Xhonneux, Louis-Pascal and Qu, Meng and Tang, Jian},
  year = {2022},
  primaryclass = {cs.LG},
  publisher = {arXiv},
  doi = {10.48550/arXiv.2202.08320}
}

@inproceedings{pyg,
  title = {Fast Graph Representation Learning with PyTorch Geometric},
  author = {Fey, Matthias and Lenssen, Jan E.},
  booktitle = {ICLR Workshop on Representation Learning on Graphs and Manifolds},
  year = {2019},
  location = {New Orleans, LA, USA},
  primaryclass = {cs.LG},
  doi = {10.48550/arXiv.1903.02428}
}

@misc{dgl,
  title = {Deep Graph Library: A Graph-Centric, Highly-Performant Package for Graph Neural Networks},
  author = {Wang, Minjie and Zheng, Da and Ye, Zihao and Gan, Quan and Li, Mufei and Song, Xiang and Zhou, Jinjing and Ma, Chao and Yu, Lingfan and Gai, Yu and Xiao, Tianjun and He, Tong and Karypis, George and Li, Jinyang and Zhang, Zheng},
  year = {2019},
  primaryclass = {cs.LG},
  publisher = {arXiv},
  doi = {10.48550/arXiv.1909.01315}
}

@inproceedings{liu2021swin,
  title={Swin transformer: Hierarchical vision transformer using shifted windows},
  author={Liu, Ze and Lin, Yutong and Cao, Yue and Hu, Han and Wei, Yixuan and Zhang, Zheng and Lin, Stephen and Guo, Baining},
  booktitle={Proceedings of the IEEE/CVF International Conference on Computer Vision (ICCV)},
  pages={10012--10022},
  year={2021},
  publisher={IEEE},
  location={Montreal, QC, Canada},
  url={https://openaccess.thecvf.com/content/ICCV2021/html/Liu_Swin_Transformer_Hierarchical_Vision_Transformer_Using_Shifted_Windows_ICCV_2021_paper.html}
}

@article{falldar,
  title={Rethinking Fall Detection With Wi-Fi},
  author={Zheng Yang and Yi Zhang and Qian Zhang},
  journal={IEEE Transactions on Mobile Computing},
  year={2023},
  volume={22},
  pages={6126-6143},
  url={https://api.semanticscholar.org/CorpusID:261435299}
}

@inproceedings{atherocsi,
     author = {Xie, Yaxiong and Li, Zhenjiang and Li, Mo},
     title = {Precise Power Delay Profiling with Commodity {WiFi}},
     booktitle = {Proceedings of the 21st Annual International Conference on Mobile Computing and Networking},
     series = {MobiCom '15},
     year = {2015},
     isbn = {978-1-4503-3619-2},
     location = {Paris, France},
     pages = {53--64},
     numpages = {12},
     url = {https://doi.org/10.1145/2789168.2790124},
     doi = {10.1145/2789168.2790124},
     acmid = {2790124},
     publisher = {Association for Computing Machinery},
     address = {New York, NY, USA},
}

@article{signfi,
author = {Ma, Yongsen and Zhou, Gang and Wang, Shuangquan and Zhao, Hongyang and Jung, Woosub},
title = {{SignFi}: Sign Language Recognition Using {WiFi}},
year = {2018},
issue_date = {March 2018},
publisher = {Association for Computing Machinery},
address = {New York, NY, USA},
volume = {2},
number = {1},
url = {https://doi.org/10.1145/3191755},
doi = {10.1145/3191755},
journal = {Proceedings of the ACM on Interactive, Mobile, Wearable and Ubiquitous Technologies},
month = mar,
articleno = {23},
numpages = {21}
}

@inproceedings{nexmon,
  author    = {Francesco Gringoli and Matthias Schulz and Jakob Link and Matthias Hollick},
  title     = {Free Your {CSI}: A Channel State Information Extraction Platform for Modern {Wi-Fi} Chipsets},
  booktitle = {Proceedings of the 13th International Workshop on Wireless Network Testbeds, Experimental Evaluation \& Characterization},
  series    = {WiNTECH '19},
  year      = {2019},
  location  = {Los Cabos, Mexico},
  pages     = {21--28},
  publisher = {Association for Computing Machinery},
  address   = {New York, NY, USA},
  doi       = {10.1145/3349623.3355477},
  url       = {https://doi.org/10.1145/3349623.3355477}
}

@inproceedings{esp32,
  author    = {Steven M. Hernandez and Eyuphan Bulut},
  title     = {Performing {WiFi} Sensing with Off-the-shelf Smartphones},
  booktitle = {2020 IEEE International Conference on Pervasive Computing and Communications Workshops (PerCom Workshops)},
  pages     = {1--3},
  publisher = {IEEE},
  location  = {Austin, TX, USA},
  doi       = {10.1109/PERCOMWORKSHOPS48775.2020.9156194},
  url       = {https://doi.org/10.1109/PERCOMWORKSHOPS48775.2020.9156194},
  month     = mar,
  year      = {2020}
}

@manual{dca1000evm,
  title        = {DCA1000EVM Data Capture Card User's Guide},
  author       = {{Texas Instruments}},
  organization = {Texas Instruments},
  number       = {SPRUIJ4A},
  year         = {2019},
  month        = may,
  note         = {Rev. A},
  url          = {https://www.ti.com/lit/pdf/spruij4}
}

@inproceedings{zhangGaitIDRobustWiFi2020,
  title = {{{GaitID}}: {{Robust Wi-Fi Based Gait Recognition}}},
  shorttitle = {{{GaitID}}},
  booktitle = {Wireless {{Algorithms}}, {{Systems}}, and {{Applications}} ({{WASA}} 2020)},
  author = {Zhang, Yi and Zheng, Yue and Zhang, Guidong and Qian, Kun and Qian, Chen and Yang, Zheng},
  editor = {Yu, Dongxiao and Dressler, Falko and Yu, Jiguo},
  year = {2020},
  series = {Lecture {{Notes}} in {{Computer Science}}},
  pages = {730--742},
  publisher = {Springer International Publishing},
  address = {Cham},
  location = {Qingdao, China},
  doi = {10.1007/978-3-030-59016-1_60},
  isbn = {978-3-030-59016-1},
  langid = {english}
}

\clearpage
\appendix

\section{Survey scope and GitHub issue analysis}
\label{app:survey-github-pipeline}

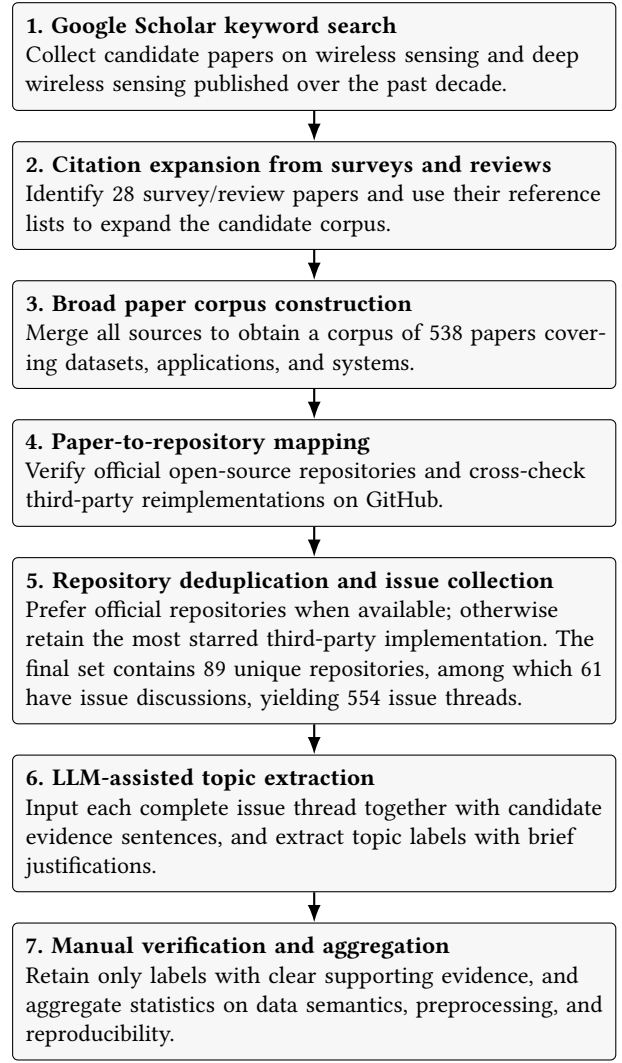
\begin{figure}[t]
  \centering
  \small
  \begin{tikzpicture}[
    node distance=4mm,
    box/.style={
      rectangle,
      rounded corners=2pt,
      draw=black,
      fill=black!3,
      text width=0.90\columnwidth,
      align=left,
      inner sep=5pt
    },
    arrow/.style={-Latex, thick}
  ]
    \node[box] (s1) {\textbf{1. Google Scholar keyword search}\\
    Collect candidate papers on wireless sensing and deep wireless sensing published over the past decade.};
    \node[box, below=of s1] (s2) {\textbf{2. Citation expansion from surveys and reviews}\\
    Identify 28 survey/review papers and use their reference lists to expand the candidate corpus.};
    \node[box, below=of s2] (s3) {\textbf{3. Broad paper corpus construction}\\
    Merge all sources to obtain a corpus of 538 papers covering datasets, applications, and systems.};
    \node[box, below=of s3] (s4) {\textbf{4. Paper-to-repository mapping}\\
    Verify official open-source repositories and cross-check third-party reimplementations on GitHub.};
    \node[box, below=of s4] (s5) {\textbf{5. Repository deduplication and issue collection}\\
    Prefer official repositories when available; otherwise retain the most starred third-party implementation. The final set contains 89 unique repositories, among which 61 have issue discussions, yielding 554 issue threads.};
    \node[box, below=of s5] (s6) {\textbf{6. LLM-assisted topic extraction}\\
    Input each complete issue thread together with candidate evidence sentences, and extract topic labels with brief justifications.};
    \node[box, below=of s6] (s7) {\textbf{7. Manual verification and aggregation}\\
    Retain only labels with clear supporting evidence, and aggregate statistics on data semantics, preprocessing, and reproducibility.};

    \draw[arrow] (s1.south) -- (s2.north);
    \draw[arrow] (s2.south) -- (s3.north);
    \draw[arrow] (s3.south) -- (s4.north);
    \draw[arrow] (s4.south) -- (s5.north);
    \draw[arrow] (s5.south) -- (s6.north);
    \draw[arrow] (s6.south) -- (s7.north);
  \end{tikzpicture}
  \caption{Survey scope definition, repository verification, and GitHub issue analysis workflow.}
  \Description{A seven-step top-to-bottom flowchart showing how the survey corpus was constructed, linked to repositories, deduplicated, analyzed with LLM assistance, and manually verified.}
  \label{fig:survey-github-flow}
\end{figure}

This appendix describes how we defined the survey scope, established the mapping between published papers and open-source repositories, and analyzed GitHub issue discussions. Our goal was not to claim exhaustive coverage of the entire wireless sensing literature, but rather to construct an auditable evidence chain linking papers, code release, and community-side reproduction discussions. \autoref{fig:survey-github-flow} provides an overview of the full procedure, and the following paragraphs explain each step in more detail.

\paragraph{Step 1: Candidate paper collection.}
We first performed keyword-based searches on Google Scholar to collect candidate papers related to wireless sensing and deep wireless sensing published over approximately the past ten years.

\paragraph{Step 2: Expansion through surveys and reviews.}
To broaden coverage beyond keyword retrieval, we additionally identified 28 survey and review papers in wireless sensing within a similar time range, and used their reference lists as an auxiliary source for citation expansion.

\paragraph{Step 3: Construction of a broad paper corpus.}
After merging the keyword-search results with the survey-based citation expansion results, we obtained a corpus of 538 papers spanning three broad categories: datasets, applications, and systems.

\paragraph{Step 4: Establishing paper-to-repository mappings.}
For each paper in the corpus, we checked whether code had been released in the open-source community and identified the corresponding GitHub repository when available. At the same time, we also searched GitHub for third-party implementations. To ensure traceability, a repository was retained only if it explicitly contained the paper title or clearly stated that it reproduced a specific published work. This enabled cross-verification from both the paper side and the repository side.

\paragraph{Step 5: Repository deduplication and issue collection.}
When multiple repositories corresponded to the same paper, we preferentially retained the official implementation released by the authors. If no official implementation was available, we retained the third-party implementation with the highest star count. This procedure yielded 89 unique repositories. Among them, 61 repositories contained issue discussions, giving a total of 554 independent issue threads for subsequent analysis.

\paragraph{Step 6: LLM-assisted topic extraction.}
We used a large language model to analyze each issue thread and extract candidate evidence sentences related to three themes central to our study: data semantics, preprocessing, and reproducibility. The LLM was used only at the candidate extraction stage rather than as the final source of labels. To make the procedure transparent, \autoref{fig:issue-prompt-card} presents key excerpts from the actual prompt template, including the constraints on evidence selection, the allowed topic categories, and the output rules.

\begin{figure}[t]
  \centering
  \small
  \begin{tikzpicture}
    \node[
      rectangle,
      rounded corners=3pt,
      draw=black,
      fill=black!3,
      text width=0.94\columnwidth,
      align=left,
      inner sep=7pt
    ] (card) {%
      \textbf{Prompt Card: Key Excerpts from the Actual Issue-Thread Template}\\[4pt]
      \texttt{Task:} ``Analyze exactly one GitHub issue thread \ldots'' and ``use the complete thread context \ldots''; ``must not classify sentences by raw keyword matching alone.''\\[4pt]
      \texttt{Key Requirement:} keep a sentence only if it ``describes a user's concrete problem, confusion, failure \ldots'' or ``provides a reply-side clarification \ldots that directly resolves or explains such a problem.''\\[4pt]
      \texttt{Category Set:} \texttt{data}, \texttt{preprocessing}, \texttt{reproducibility}, and \texttt{torchradio\_relevant}.\\[4pt]
      \texttt{Output Rules:} ``Return JSON only,'' ``cite only sentence ids that exist in the provided sentence inventory,'' and ``keep only the strongest evidence sentences.''\\[4pt]
      \texttt{Decision Style:} ``Be conservative'' and ``prefer precision over recall.''%
    };
  \end{tikzpicture}
  \caption{Key constraints excerpted from the actual prompt template used for issue-thread analysis.}
  \Description{A rounded-corner card showing the key parts of the real GitHub issue thread analysis prompt, including the task definition, evidence requirements, category set, output rules, and conservative decision style.}
  \label{fig:issue-prompt-card}
\end{figure}

\paragraph{Step 7: Manual verification and aggregation.}
We manually reviewed the LLM outputs and retained only topic labels supported by explicit evidence. The final topic counts are summarized in \autoref{tab:issue-topic-stats}.

\begin{table}[t]
  \centering
  \footnotesize
  \caption{Topic statistics from 554 GitHub issue threads after manual verification.}
  \label{tab:issue-topic-stats}
\begin{tabular}{cccc}
\hline
Topic           & \begin{tabular}[c]{@{}c@{}}No. of \\ threads\end{tabular} & \begin{tabular}[c]{@{}c@{}}No. of evidence\\ sentences\end{tabular} & \begin{tabular}[c]{@{}c@{}}No. of \\ repositories\end{tabular} \\ \hline
Data semantics  & 258                                                                 & 366                                                                 & 44                                                             \\
Preprocessing   & 102                                                                  & 153                                                                 & 31                                                             \\
Reproducibility & 65                                                                 & 99                                                                 & 32                                                             \\ 
Others & 129                                                                 & 158                                                                 & 37                                                             \\ \hline
\end{tabular}
\end{table}

These statistics indicate that community discussions repeatedly expose problems related to unclear data meaning, opaque preprocessing procedures, and missing implementation details required for reproduction. We therefore use GitHub issue threads as direct evidence of the reuse burden encountered by the community, rather than as an unbiased estimate of prevalence for the field as a whole.

\end{document}